\documentclass[12pt]{iopart}
\usepackage{iopams}
\usepackage{graphicx, subfigure}
\expandafter\let\csname equation*\endcsname\relax
\expandafter\let\csname endequation*\endcsname\relax
\usepackage{amsmath}
\usepackage[utf8]{inputenc}
\usepackage{color}
\usepackage{bm}
\usepackage{cite}
\usepackage{url} 
\newtheorem{theorem}{Theorem}
\newtheorem{proposition}[theorem]{Proposition}

\newcommand{\Dirac}{{\boldsymbol{\mathcal{D}}} \hskip -2.5mm \slash}
\definecolor{brickred}{rgb}{0.7, 0.25, 0.33}
\definecolor{applegreen}{rgb}{0.55, 0.71, 0.0}

\def\be{\begin{equation}}
\def\ee{\end{equation}}
\def\bc{\begin{center}}
\def\ec{\end{center}}
\def\bea{\begin{eqnarray}}
\def\eea{\end{eqnarray}}

\newcommand\correspondingauthor{\footnote{Corresponding author.}}
\begin{document}

\title{Global Topological Dirac Synchronization}

\author{Timoteo Carletti$^1$, Lorenzo Giambagli$^2$, Riccardo Muolo$^3$, Ginestra Bianconi$^{4}$\correspondingauthor}
\address{$^1$Department of Mathematics \& naXys, Namur Institute for Complex Systems, University of Namur, Rue Grafé 2, B5000 Namur, Belgium\\
$^2$ Department of Physics, Freie Universi\"at Berlin, Arnimallee 12, 14195, Berlin, Germany\\
$^3$ Department of Systems and Control Engineering, Institute of Science Tokyo (former Tokyo Tech), O-okayama 2-12-1, Meguro, Tokyo 152-8552, Japan\\
$^4$ Centre for Complex Systems, School of Mathematical Sciences, Queen Mary University of London, London, E1 4NS, United Kingdom}

\ead{ginestra.bianconi@gmail.com}
\begin{abstract}
Synchronization is a fundamental dynamical state of interacting oscillators, observed, {e.g.}, in natural biological rhythms and in the brain.  Global synchronization which occurs when non-linear or chaotic oscillators placed on the nodes of a network display the same dynamics has received great attention in network theory. Here we propose and investigate Global Topological Dirac Synchronization on higher-order networks such as cell and simplicial complexes. This is a state where oscillators associated to simplices and cells of arbitrary dimension, coupled by the Topological Dirac operator, operate at unison. By combining algebraic topology with non-linear dynamics and machine learning,  we derive the topological conditions under which this state exists and the dynamical conditions under which it is stable. We provide evidence of $1$-dimensional simplicial complexes (networks) and $2$-dimensional simplicial and  cell complexes where Global Topological Dirac Synchronization can be observed. Our results point out that Global Topological Dirac Synchronization is a possible  dynamical state of  cell complexes  and simplicial complexes that occur only in some specific network topologies and geometries, the latter ones being determined by the weights of the higher-order networks. 
\end{abstract}


\section{Introduction}

Global synchronization, the spontaneous ability of coupled {identical} oscillators to operate at unison and thus exhibit a coherent collective behavior, is a widespread phenomenon at the root of several biological rhythms or human made technological systems~\cite{Pikovsky2001,arenasreview}.  When  coupled identical oscillators are associated to the nodes of the network, the global synchronization state is guaranteed to always exist, but its stability is determined by the Master Stability Function  initially proposed by Fujiska and Yamada~\cite{fujisaka1983stability} and then reformulated  by Pecora and Carroll~\cite{Pecora,Pecora_etal97} and Barahona and Pecora on small-world networks ~\cite{barahona2002synchronization}.

 Recently synchronization on higher-order networks~\cite{bianconi2021higher,millan2025topology,bick2023higher,battiston2020networks,battiston2021physics,majhi2022dynamics} where the interactions among the oscillators are many-body, is attracting great attention~\cite{millan2020explosive,ghorbanchian2021higher,carletti2023global,WMCB2024,skardal2019abrupt,tanaka2011multistable,leon2024higher,lucas2020multiorder,Krawiecki2014,carletti2020dynamical,gambuzza2021stability,gallo2022synchronization,MKJ2020}. Higher-order networks~ \cite{bianconi2021higher,bick2023higher,battiston2020networks,battiston2021physics,majhi2022dynamics} (such as hypergraphs, simplicial and cell complexes)  capture the function of many complex systems, e.g., brain networks~\cite{giusti2016two,reimann2017cliques}, social networks~\cite{patania2017shape} and protein interaction networks~\cite{estradaJTB}. 
Dynamics on higher-order networks~\cite{millan2025topology} including synchronization~\cite{millan2020explosive,ghorbanchian2021higher,carletti2023global,WMCB2024,skardal2019abrupt,tanaka2011multistable,leon2024higher,lucas2020multiorder,Krawiecki2014,carletti2020dynamical,gambuzza2021stability,gallo2022synchronization,MKJ2020}, random walks \cite{schaub2020random,carletti2020random}, pattern formation \cite{carletti2020dynamical,muolo2023turing}, and higher-order diffusion \cite{torres2020simplicial,ziegler2022balanced}, often displays phenomena that have no equivalent on simple networks~\cite{battiston2021physics,majhi2022dynamics,bianconi2021higher}.

Of special relevance here is the great progress recently made in unveiling the interplay between topology and dynamics of higher-order networks~\cite{millan2025topology}. This new emergent field reveals new collective dynamical states  of  Topological Synchronization~\cite{millan2020explosive,ghorbanchian2021higher,carletti2023global,WMCB2024}, captured by the Topological Kuramoto model~\cite{millan2020explosive,ghorbanchian2021higher} and the Global Topological synchronization \cite{carletti2023global,WMCB2024}.  
These two novel classes of synchronization models defined on simplicial and cell complexes describe collective phenomena of  {\em topological signals}, i.e., dynamical variables associated  not only to nodes, but also to links, triangles and higher-dimensional simplices or cells. Examples of real topological signals are edge signals, such as  synaptic and brain edge signals~\cite{faskowitz2022edges}, biological transportation fluxes or traffic signals~\cite{sardellitti2021topological}, or climate data, such as currents in the ocean or speed of wind at different locations~\cite{schaub2021signal,calmon2023dirac}. As such, topological signals are at the forefront of Topological Machine Learning and Signal Processing~\cite{sardellitti2021topological,schaub2021signal,calmon2023dirac,wang2024dirac,bodnar2021weisfeiler,ebli2020simplicial}. 

Topological Synchronization demonstrates on one side how dynamics can learn topology and how topology can shape dynamics~\cite{millan2025topology,millan2020explosive,ghorbanchian2021higher,carletti2023global,WMCB2024}. In particular, higher-order Topological Synchronization described by the Topological Kuramoto~\cite{millan2020explosive,ghorbanchian2021higher,calmon2021topological,arnaudon2022connecting} or by the Global Topological Synchronization~\cite{carletti2023global,WMCB2024} can be shown to localize on the higher-dimensional holes of the simplicial and cell complex, showing how this dynamics can reveal the underlying topology of the simplicial or cell complex over which it is defined.  

 Global Topological Synchronization (GTS)~\cite{carletti2023global} refers to the dynamical state of $k$-dimensional topological signals defined on edges ($k=1$), triangles or squares ($k=2$), or higher dimensional simplices and cells, in which all the identical oscillators supporting the $k$-topological signals display the same dynamics when they are coupled together via the $k$-th Hodge Laplace operator. Interestingly, Global Topological Synchronization has very distinct properties with respect to global synchronization of node signals. Indeed, while for identical oscillators associated exclusively to the nodes of the network the globally synchronized state always exists but might not be dynamically stable, for the GTS the synchronous state exists only for simplicial complexes obeying specific conditions on the spectrum of their Hodge Laplacian~\cite{carletti2023global}. While  GTS might not be in general guaranteed, on one side there are some topologies, like the square lattice tessellation of the $K$ dimensional torus ($K$ dimensional lattice with periodic boundary conditions), that allow GTS of the topological signals of every dimension, on the other side, by considering weighted version of the simplicial and cell complexes can allow GTS also if the unweighted structure impedes it \cite{WMCB2024}.

 In all these works, GTS has been studied by considering topological signals of the same dimension $k$. However, the same higher-order networks can sustain dynamical signals of different dimension at the same time, and thus it is an interesting natural question to investigate their coupled dynamics.
 The key topological operator that couples the dynamics of the topological signals of different dimensions is the {\em Dirac operator} \cite{bianconi2021topological,lloyd2016quantum}. This operator was  originally proposed in the framework of lattice gauge theory~\cite{kogut1975hamiltonian,becher1982dirac}, and continues to inspire works in theoretical physics \cite{bianconi2023mass,delporte2023dirac,bianconi2024quantum}. However, only recently it has demonstrated its pivotal role in network science and machine learning~\cite{bianconi2021topological,lloyd2016quantum} and it has been adopted in the study of nonlinear dynamics~\cite{calmon2021topological,calmon2023local,nurisso2024unified}, pattern formation~\cite{giambagli2022diffusion,muolo2024three} signal processing~\cite{calmon2023dirac}, topological neural networks~\cite{nauck2024dirac}, and quantum persistence homology~\cite{lloyd2016quantum,wee2023persistent,suwayyid2024persistent,suwayyid2024persistent2,ameneyro2022quantum}.

{  In this work, we combine advanced concepts of algebraic topology and the latest developments of nonlinear research and machine learning to provide evidence of Global Topological Dirac Synchronization (GTDS).
 Global Topological Dirac Synchronization is a dynamical state of identical oscillators defined on nodes, edges, triangles, and, in general, on every $k$-dimensional cell of a higher-order network, whose topological signals obey the same dynamics. 
 In this model, the topological signals of different dimension are coupled via the Dirac operator and its associated gamma matrices, that here play the role of higher-order coupling constants.

It is well known that  global synchronization on a graph, i.e., the dynamical state in which all the nodes have the same dynamics, always exists, for any arbitrary graph and identical oscillators. Thus, in the context of network theory the research has been focusing  exclusively on the characterization of the stability of such a dynamical state. On the contrary, here we show that  GTDS is a dynamical state that can be observed only on specific topologies and we provide the most general conditions under which a simplicial or cell complex might admit this dynamical state and the general stability criteria of these states.

We demonstrate that  Eulerian graphs can admit GTDS if the dynamics is defined on nodes and edges, and we combine machine-learning to nonlinear dynamics to predict the regions of stability of the GTDS in parameter space. 
Moving from graph to higher-order networks,  we provide constructive proofs that some cell complexes such as any $K$-dimensional torus with hypercube tessellations, can admit GTDS for any arbitrary dimension $K$.
Finally, we show that GTDS on simplicial complexes can be observed only by attributing to simplices weights encoded in their associated metric matrices with these weights obeying specific algebraic conditions.

The proposed theoretical framework greatly extends the global synchronization model which is known to have a large variety of applications, from  biological rhythms or human made technological systems~\cite{Pikovsky2001,arenasreview}. Moreover, the proposed model of GTDS has the potential to relate to the growing interest on synchronization states in condensed matter systems ranging from nano and mechanical oscillators to interacting electron systems \cite{albertsson2021ultrafast,matheny2019exotic,flovik2016describing,nag2019dynamical,delmonte2023quantum}. In particular, we believe that GTDS could inspire new experiments with nano-oscillators that are already attracting increasing attention as important new technologies for computing \cite{albertsson2021ultrafast} and that future research could investigate the possible experimental realization of this model in the lab.
}

\section{Simplicial and cell complexes}
\label{sec:sccomplexes}

Higher-order networks  \cite{bianconi2021higher} are generalized network structures composed by nodes and edges, but also triangles, tetrahedra and higher-order structures that encode the many-body interactions in complex systems. Here we focus in particular on simplicial complexes (see Figure $\ref{fig:sketch}$) and on cell complexes. A {\em  $k$-simplex}, $\sigma^{k}$, is a set of $k+1$ nodes. The simplices that are formed by a proper subset of the nodes of a $k$-simplex are called its {\em faces}. Two simplices are incident if and only if either they share a common face or one is the face of the other. A simplicial complex $\mathcal{X}$  is a set of simplices closed under the inclusion of faces, namely if $\sigma \in \mathcal{X}$, then also all the faces of $\sigma$ should belong to $\mathcal{X}$.  
Cell complexes generalize simplicial complexes and they are defined as a set of cells (or regular polytopes) closed under the inclusions of the faces of the polytopes. Thus cell complexes are built from simplices, hypercubes, orthoplexes etc. and notably they include important topologies such as the square lattice tessellation of $2D$-torus and cubic tessellation of a $3D$-torus. Here and in the following we will indicate with $\sigma^k_i$ the $i$-th $k$-dimensional simplex (cell) of the simplicial complex (cell complex) and with $K$ the dimension of the simplicial (cell) complex given by the largest dimension of its simplices (cells).
Moreover, we denote by $N_k$, $k=0,\dots, K$, the number of $k$-cell simplices (cells) in the simplicial (cell) complex.

Until now we have discussed exclusively unweighted simplicial and cell complexes.
However simplicial or cell complexes can be associated to metric matrices if the simplices or cells, $\sigma_i^k,$ are assigned a weight $w_i^k>0$ that can be interpreted as an {\em affinity weight}~\cite{GradyPolimeni2010}.
In this case the metric matrices ${\bf G}_k$ are $N_k\times N_k$ diagonal matrices of non-zero elements 
\begin{equation}
{G}_{k}(i,i)=\frac{1}{w_{i}^k}\, .
\label{G}
\end{equation}
As we will see in the next paragraphs, these metric matrices play a central role to define how the weights modify the exterior calculus operations such as the gradient, the divergence and the curl.

\section{The dynamical state of a higher-order network}
\begin{figure}[h]
\centering
\includegraphics[width=0.95\textwidth]{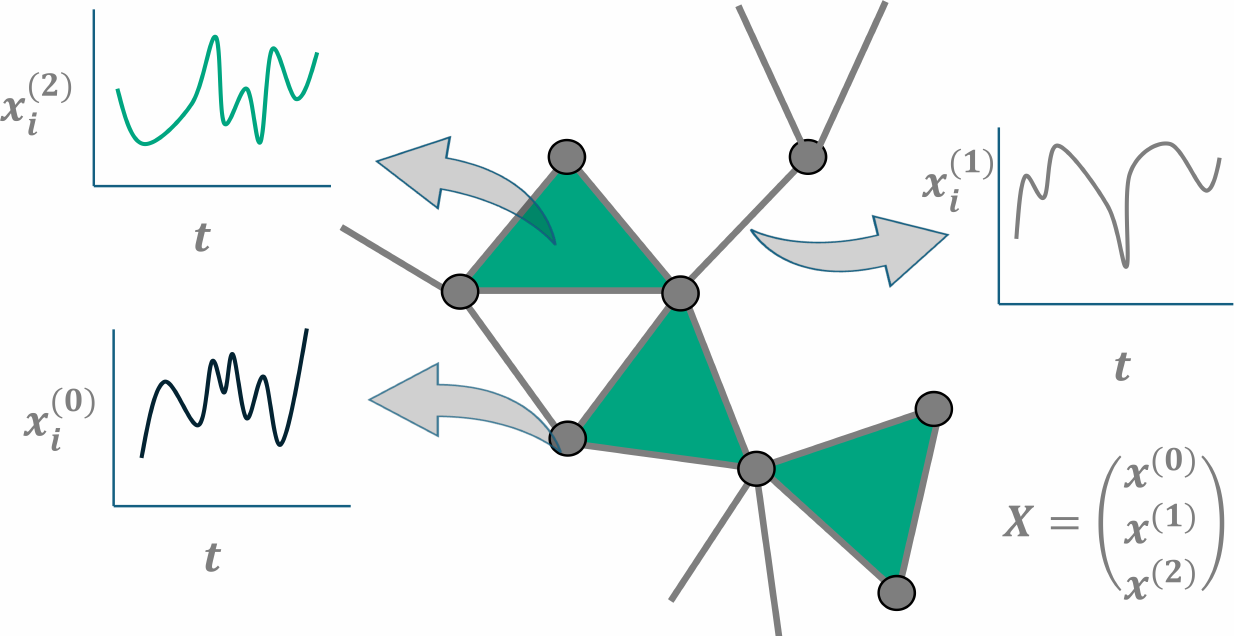}
\caption{The dynamical state of a simplicial complex is encoded in the topological spinor $\mathbf{X}$ given by the direct sum of the topological signals of different dimensions. Thus, the dynamical state of a simplicial complex of dimension $K=2$ (shown in the Figure) is encoded in the topological spinor $\mathcal{X}=(\mathbf{x^{(0)}},\mathbf{x^{(1)}},\mathbf{x^{(2)}})^{\top}$, where $\mathbf{x^{(0)}},$ $\mathbf{x^{(1)}},$ and $\mathbf{x^{(2)}}$ indicate the node signals, the  edge signals and the triangle signals, respectively, of the simplicial complex.}
\label{fig:sketch}
\end{figure}

Recently, it has been realised that, when considering the dynamical state of a higher-order network, it may be beneficial to abandon the node centred point of view that associates dynamical variables only to the nodes of the higher-order networks, and to consider, instead, topological signals. The latter are dynamical variables that can be associated not only to the nodes of the higher-order network, but also to the edges, triangles and higher-dimensional simplices and cells of the considered structure.
In order to fully describe the dynamical state of the network, we need to consider the {\em topological spinor} ${\bf X}$ \cite{bianconi2021topological} (see Figure $\ref{fig:sketch}$), which is given by the direct sum of the topological spinors associated to each simplex (cell) of the higher-order network.

Without loss of generality, here we consider simplicial or cell complexes of dimension $K=2$, i.e., formed by nodes, edges, and triangles (or $2D$ polygons), and we indicate with $\mathcal{N}=N_0+N_1+N_2$ the {total number} of all simplices (cells) of the complex. We define the topological spinor ${\bf X}$ as the $\mathcal{N}$ column vector  
\be
\mathbf{X}=\begin{pmatrix}{\mathbf x}^{(0)}\\{\mathbf x}^{(1)}\\{\mathbf x}^{(2)}\\\end{pmatrix},
\label{ts}
\ee
where ${\mathbf x}^{(0)},{\mathbf x}^{(1)}$ and ${\mathbf x}^{(2)}$ indicate the node, the edge and the triangle (polygon) signals, which are given by
\be
{{\bf x}^{(0)}}=\begin{pmatrix}
			{{x}^{(0)}_1}\\
			{{x}^{(0)}_2}\\
			\vdots\\
			{{x}^{(0)}_i}\\
			\vdots\\
			{{x}^{(0)}_{N_0}}
		\end{pmatrix}\quad {{\bf x}^{(1)}}=\begin{pmatrix}
			{{x}^{(1)}_1}\\
			{{x}^{(1)}_2}\\
			\vdots\\
			{{x}^{(1)}_i}\\
			\vdots\\
			{{x}^{(1)}_{N_1}}\end{pmatrix}\quad {\bf x}^{(2)}=\begin{pmatrix}
			{{x}^{(2)}_1}\\
			{{x}^{(2)}_2}\\
			\vdots\\
			{{x}^{(2)}_i}\\
			\vdots\\
			{{x}^{(2)}_{N_2}}\end{pmatrix}\, ,	
   \label{ts2}
\ee
where we denoted by $x^{(k)}_i$ the dynamical variable(s)associated to the $i$-th $k$-dimensional simplex (cell) of the higher-order network.
Recent results \textcolor{black}{have} shown that topological signals can undergo collective phenomena \cite{millan2020explosive,carletti2023global,WMCB2024,calmon2021topological,ghorbanchian2021higher,giambagli2022diffusion,muolo2024three}. It particular, in \cite{carletti2023global,WMCB2024} it has been shown that under suitable topological, geometrical and dynamical conditions, the $k$-dimensional topological signal can undergo Global Topological Synchronization (GTS). However, an important open question so far is whether the whole topological spinor, involving all the topological signals of the higher-order network, can undergo Global Topological Synchronization transition when the topological spinors of different dimensions are coupled to each other. This is the main question we will answer with the present work.

\section{Basics of exterior calculus}
In order to define dynamical processes acting on the topological signals $\mathbf{x}^{(k)}$, we introduce some key exterior calculus operators that are instrumental to define fundamental discrete operators such as the discrete gradient, discrete divergence and discrete curl on simplicial and cell complexes. These  discrete calculus operators can be defined thanks to algebraic topology in terms of rectangular matrices called {\em boundary matrices}. These matrices are also fundamental to define Hodge Laplacians that describe higher-order diffusion on simplicial and cell complexes. 

\subsection{Unweighted boundary operator}
The boundary operators are rooted in algebraic topology and play a fundamental role in exterior calculus.
In algebraic topology~\cite{bianconi2021topological}, cells are assigned an orientation, {  here,  we assume typically that  the orientation is induced by the node labels. However, for the $K$-dimensional torus we consider the usual orientation that guarantees periodic boundary conditions.} A coherent orientation of a $(k-1)$-face $\sigma^{k-1}$ of a $k$-cell $\sigma^k$, will be denoted by $\sigma^{k-1}\sim \sigma^{k}$, otherwise we will write $\sigma^{k-1}\not\sim \sigma^{k}$. 
A simplicial or cell complex can be encoded via the set of its boundary matrices $\mathbf{B}_k$. Each unweighted boundary matrix $\mathbf{B}_k^{(U)}$ is a $N_{k-1}\times N_k$  matrix of elements 
\begin{equation}
\label{eq:Bk}
{B}_k^{(U)}(i,j)=
\begin{cases}
 1 & \text{ if } \sigma_i^{k-1}\sim \sigma_j^{k},\\
 -1 & \text{ if } \sigma_i^{k-1}\not\sim \sigma_j^{k},\\
  0 & \text{ otherwise}\, ,
\end{cases}
\end{equation}
 for all $k=1,\dots,K$, $K$ being the dimension of the cell complex. From this definition it is possible to prove the fundamental topological property stated as {\em the boundary of the boundary is null}, which implies
 \be
 \mathbf{B}_k^{(U)}\mathbf{B}_{k+1}^{(U)}=0\, ,
 \label{boundary_unweighted}
 \ee
 that characterizes boundary operators. 

\subsection{Metric boundary operators}
For weighted simplicial and cell complexes \cite{baccini2022weighted,WMCB2024}, the definition of the {\em weighted boundary matrices}, ${\mathbf{B}}_{k}$, is strongly affected by the metric matrices ${\bf G}_k$ defined in Eq.~\eqref{G}, and can be defined as 
\begin{equation}
{\mathbf{B}}_{k}=\mathbf{G}_{k-1}^{1/2}\mathbf{B}_{k}^{(U)}\mathbf{G}_{k}^{-1/2}\, ,
\label{eq:Bweigthed}
\end{equation}
where $\mathbf{B}_{k}^{(U)}$ is the unweighted boundary matrix defined in Eq.~\eqref{eq:Bk}.
Interestingly, it can be easily proved that, starting from the definition of the weighted boundary operator and from Eq.~\eqref{boundary_unweighted}, the weighted boundary matrices defined in this way also satisfy the topological relation that the {\em boundary of the boundary is null} 
 \be
 \mathbf{B}_k\mathbf{B}_{k+1}=0\, .
 \ee
The unweighted complexes can be recovered by assuming all the metric matrices to be trivial, i.e., given by the identity matrix ${\bf G}_k={\bf I}_{N_k}$; indeed in this case we get $\mathbf{B}_{k}=\mathbf{B}_{k}^{(U)}$. Justified by this consideration, in the following we will denote by $\mathbf{B}_{k}$  both unweighted and weighted boundary matrices indicating, when necessary, whether we consider unweighted or weighted complexes.

 The boundary matrix ${\bf B}_1^{\top}$ acts on the node topological signal $\mathbf{x}^{(0)}$ as the discrete gradient, the boundary matrix ${\bf B}_2^{\top}$ acts on the edge topological signal $\mathbf{x}^{(1)}$ as the discrete curl, while the boundary matrix ${\bf B}_1$ acts on the edge topological signal $\mathbf{x}^{(1)}$ as the discrete divergence.
 
\subsection{The Hodge Laplacians}
The Hodge Laplacian  $\mathbf{L}_k$~\cite{bianconi2021higher,eckmann1944harmonische,torres2020simplicial,GradyPolimeni2010,lim2020hodge,
horak2013spectra} describes diffusion for $k$-cells to $k$-cells going either through a $(k-1)$-cell or a $(k+1)$-cell. This linear operator can be encoded in a $N_k\times N_k$ matrix defined as:
\begin{equation}
\label{eq:L}
\mathbf{L}_k = \mathbf{L}_k^{\mathrm{up}}+\mathbf{L}_k^{\mathrm{down}} \, , \quad k=1,\dots, K-1\, ,
\end{equation}
where $\mathbf{L}_k^{\mathrm{up}}=\mathbf{B}_{k+1}\mathbf{B}_{k+1}^\top$ denotes the upper Laplace matrix and $\mathbf{L}_k^{\mathrm{down}}=\mathbf{B}_k^\top\mathbf{B}_{k}$ the lower Laplace matrix. Let us observe that for $k=0$ and $k=K$ those definitions reduce to $\mathbf{L}_0 = \mathbf{B}_1\mathbf{B}_{1}^\top$, and $\mathbf{L}_K = \mathbf{B}_K^\top\mathbf{B}_{K}$ as $\mathbf{B}_{K+1}=\mathbf{0}$, and we have, by definition, $\mathbf{B}_{0}=\mathbf{0}$ as well. For $k=0$, the Hodge Laplacian  $\mathbf{L}_{0}$ coincides with the (combinatorial) graph Laplacian of the network. 
From the latter definitions one can conclude that the non-zero spectrum of $\mathbf{L}_{k}^{\textrm{down}}$ coincides with the non-zero spectrum of $\mathbf{L}_{k}^{\textrm{up}}$. Moreover, the  Hodge Laplacians satisfy the Hodge decomposition. In fact, we have 
\begin{equation}
\mathbf{L}_{k}^{\textrm{up}}\mathbf{L}_{k}^{\textrm{down}}=0 \text{ and }\mathbf{L}_{k}^{\textrm{down}}\mathbf{L}_{k}^{\textrm{up}}=0\, .
\end{equation}
We note here that we will adopt the same notation $\mathbf{L}_k$ for both weighted and unweighted Hodge Laplacians with the latter obtained by using the definition of the  boundary operator given in Eq.~\eqref{eq:Bweigthed} and by setting all the metric matrices equal to the identity, i.e., ${\bf G}_k={\bf I}_{N_k}$. When the metric matrices are non-trivial, the obtained Hodge Laplacians are weighted and symmetric reducing to the well studied~\cite{Chung} symmetric graph Laplacian for $k=0$.

The Hodge Laplacian $\mathbf{L}_k$ plays a key role in the topological dynamics of higher-order networks, but has an important limitation because it allows to exclusively deal with topological signals of the same dimension $k$. Therefore, the Hodge Laplacians do not allow for cross-talk of the topological signals of different dimensions. To go beyond this limitation, we need to consider the Topological Dirac operator, that we will introduce in details in the next Section.

\section{The Topological Dirac operator}
The Topological Dirac operator $\mathbf{D}$ is key for coupling topological signals of different dimensions and has a fundamental role in capturing the dynamics of higher-order networks. For these reasons, it is increasingly recognised in the context of network theory and machine learning. Indeed, the Dirac operator has been proposed for determining pattern formation of topological signals \cite{giambagli2022diffusion,muolo2024three,muolo2024review}, for determining explosive transitions in higher-order Kuramoto model coupled by the Dirac operator \cite{calmon2023local}, for performing signal processing of coupled topological signals \cite{calmon2023dirac} and for the formulation of novel topological neural networks \cite{nauck2024dirac}. 

On a $K=2$ dimensional simplicial (cell) complex, the Dirac operator ${\bf D}$ is a $\mathcal{N}\times\mathcal{N}$ matrix that acts on the topological spinor $\bf{{X}}$ defined in Eq.~\eqref{ts} and couples topological signals of different dimensions. Specifically, the Dirac operator $\mathbf{D}$ is defined in terms of the boundary operators as
	\be
	{\bf D}=\begin{pmatrix}
			0 & {\bf B}_{1} & 0\\
			{\bf B}_{1}^{\top }  & 0 & {\bf B}_{2} \\
			0 & {\bf B}_{2}^{\top }  & 0
		\end{pmatrix}\, .
	\ee
	Hence, the Dirac operator of a $2$-dimensional simplicial (cell) complex projects the signals of the nodes into the edges, the signal of the edges into the nodes and into the triangles (polygons), and the signal on the triangles (polygons) into the edges.
	One of the most important properties of the Dirac operator is that it can be interpreted as the ``square-root" of the higher-order Laplacian, indeed 
	\be{\bf D}^2=\boldsymbol{\mathcal{L}}=\begin{pmatrix}
			 {\bf L}_{0} & 0 &0\\
			0&{\bf L}_{1} & 0 \\
			0 & 0& {\bf L}_{2}
		\end{pmatrix}\, .
  \ee
		Note that, {  ${\bf L}_{[0]}$ and ${\bf L}_{[1]}$ have the same non-zero eigenvalues, and, similarly,  also  ${\bf L}_{[1]}^{down}$ and ${\bf L}_{[2]}^{down}$ share the same non-zero spectrum}. Moreover, since we are considering a $2$-dimensional simplicial complex, ${\bf L}_{1}={\bf L}_{1}^{\textrm{up}}+{\bf L}_{1}^{\textrm{down}}$, but ${\bf L}_{2}={\bf L}_{2}^{\textrm{down}}$.
		{  From  these properties} it follows that the eigenvalues of the Dirac operator are the square roots of the eigenvalues of the Hodge Laplacian taken both with positive and negative sign. Therefore, while the Hodge Laplacians are semi-definite positive, the Dirac operator is not.
		
	On a $K=2$ dimensional simplicial or cell complex, the Dirac operator can be expressed as the sum of two Dirac operators; the first one, ${\bf D}_{[1]}$, only acting on nodes and edges, while the second ${\bf D}_{[2]}$ only acts on edges and polygons, i.e.,
	\be
	{\bf D}={\bf D}_{[1]}+{\bf D}_{[2]}\, .
	\ee
	with 
\be	
 {\bf D}_{[1]}=\begin{pmatrix}
			0 & {\bf B}_{1} & 0\\
			{\bf B}_{1}^{\top } & 0 & 0\\
			0 & 0 & 0
		\end{pmatrix}\, ,\quad {\bf D}_{[2]}=\begin{pmatrix}
			0 & 0 & 0\\
			0 & 0 & {\bf B}_{2}\\
			0 & {\bf B}_{2}^{\top } & 0
		\end{pmatrix}\, ,
		\ee
	whose square is given by 	
		\be	
	 {\bf D}_{[1]}^2={\boldsymbol{\mathcal{L}}}_{[1]}=\begin{pmatrix}
		{\bf L}_{0} & 0 &0\\
			0&{\bf L}_{1}^{\textrm{down}} & 0 \\
			0 & 0& 0
		\end{pmatrix}\, ,\quad {\bf D}_{[2]}^2={\boldsymbol{\mathcal{L}}}_{[2]}=\begin{pmatrix}
		0 & 0 &0\\
			0&{\bf L}_{1}^{\textrm{up}} & 0 \\
			0 & 0& {\bf L}_{2}^{\textrm{down}}.
		\end{pmatrix}
		\ee
		From the definition of ${\bf D}_{[k]}$, it is immediate to check that  
		\be
		{\bf D}_{[1]}{\bf D}_{[2]}={\bf D}_{[2]}{\bf D}_{[1]}={ 0}\, ,
		\ee
		hence, the Dirac operator obeys the Dirac decomposition \cite{calmon2023dirac}, namely,
		\be
		\mbox{im}({\bf D}_{[1]})\subseteq \mbox{ker}({\bf D}_{[2]})\, ,\quad
		\mbox{im}({\bf D}_{[2]})\subseteq \mbox{ker}({\bf D}_{[1]})\, .
		\ee
		This implies that, for every topological spinor 
		$\bf X$, there is a unique way to decompose it as 
			\be
	{\bf X}={\bf X}_{[1]}+{\bf X}_{[2]}+{\bf X}^{\textrm{harm}}\, .
 \label{Diracdec}
	\ee
	Here, ${\bf X}_{[1]}$ and ${\bf X}_{[2]}$ are in the image of ${\mathbf{D}}_{[1]}$ and $\mathbf{D}_{[2]}$, respectively, and, thus, can be obtained as 
	\be
	{\bf X}_{[k]}={\bf P}_{[k]}{\bf X},
	\ee
	with ${\bf P}_{[k]}$ indicating the  projectors
	\be{\bf P}_{[k]}={\bf D}_{[k]}{\bf D}_{[k]}^+={\boldsymbol{\mathcal{L}}}_{[k]}{\boldsymbol{\mathcal{L}}}_{[k]}^{+},\label{projectors}\ee
	where ${\bf D}_{[k]}^+$ and $\boldsymbol{\mathcal{L}}_{[k]}^+$ indicate the pseudo-inverse of the Dirac and the Hodge Laplacian, respectively.
	The spinor ${\bf X}_{[1]}$ is non-zero  only on nodes and edges and its edge elements include only the irrotational component of the edge signals. The spinor ${\bf X}_{[2]}$ is non-zero only on edges and $2$-dimensional cells (triangles, squares, etc.) and its edge elements include only the solenoidal component of the edge signals. 
	Moreover, ${\bf X}^{\textrm{harm}}$ is the harmonic component of the topological spinor and it obeys
	\be
	{\bf D}_{[k]}\hat{\bf X}^{\textrm{harm}}={ 0}\, ,
	\ee
	for every $k\in \{1,2\}$.
	From the definition of the projectors ${\bf P}_{[k]}$ it follows that ${\bf P}_{[k]}{\bf D}={\bf D}_{[k]}$.

\section{Global Topological Dirac Synchronization (GTDS)}
\subsection{Dynamical equations for GTDS}
\label{sec:dyn}

Global Topological Dirac Synchronization (GTDS) occurs when the topological spinor ${\bf X}$ obeys a global synchronized dynamics. In order to study the topological and dynamical conditions allowing for the emergence of this dynamical state, we consider topological spinor defined in Eq.~\eqref{ts} and Eq.~\eqref{ts2}, where the signal associated to the generic node ${\bf x}^{(0)}_i$, the generic edge ${\bf x}_i^{(1)}$, and the generic polygons ${\bf x}_i^{(2)}$ have the same dimension $d$. Thus, we adopt the notation $\vec{x}_i^{(0)}\in \mathbb{R}^d$, $\vec{ x}^{(1)}_i\in \mathbb{R}^d$ and $\vec{ x}^{(2)}_i\in\mathbb{R}^d$. 
The topological signals of different dimension will be coupled by the Dirac operator $\Dirac$ that is constructed starting from the operators ${\bf D}_{[k]}$ coupled with the gamma matrices $\bm\gamma_{[k]}$. { This choice is dictated and inspired by the use of gamma matrices in the Topological Dirac Equation \cite{bianconi2021topological}. In the present dynamical system context, the gamma matrices encode the coupling constants of different topological signals and  provide an additional degree of freedom to account for a larger variety of dynamical states.  Interestingly the gamma matrices can be also used to generalize the present framework along the lines defined in the context of  Dirac pattern formation~\cite{muolo2024three}, where the gamma matrices allow to  couple two node signals with a single edge signal.}
Thus, we define $\Dirac$ as
\be
\Dirac={\bm\gamma}_{[1]}{\bf D}_{[1]}+{\bm\gamma}_{[2]}{\bf D}_{[2]}\, ,
\ee
where the gamma matrices $\bm\gamma_{[k]}$ are defined as
 \be	
 {\bm \gamma}_{[1]}=\begin{pmatrix}
			{\bf I}_{N_0}\otimes\bm  \gamma_0^{(1)}  & 0 & 0\\
			0 & {\bf I}_{N_1}\otimes\bm  \gamma_1^{(1)}  & 0\\
			0 & 0 & 0
		\end{pmatrix}\, ,\quad {\bm \gamma}_{[2]}=\begin{pmatrix}
			0 & 0 & 0\\
			0 & {\bf I}_{N_1}\otimes\bm  \gamma_1^{(2)} & 0\\
			0 & 0 &  {\bf I}_{N_2}\otimes\bm \gamma_2^{(2)}
		\end{pmatrix}\, ,
		\ee
  with $\bm\gamma_k^{(n)}$ being $\mathbb{R}^d\times \mathbb{R}^d$ matrices, thus, 
  \be
  \Dirac=\begin{pmatrix}
			0 & {\bf B}_{[1]}\otimes\bm  \gamma_0^{(1)} & 0\\
			{\bf B}_{[1]}^{\top }\otimes \bm  \gamma_1^{(1)} & 0 & {\bf B}_{[2]}\otimes\bm  \gamma_1^{(2)}\\
			0 & {\bf B}_{[2]}^{\top }\otimes\bm  \gamma_2^{(2)} & 0
		\end{pmatrix}\, .
  \ee
The topological signals exhibit an internal dynamics, while signals defined on incident simplices are coupled locally by the Dirac operator $\Dirac$.
Specifically, we consider the following dynamics for Global Topological Dirac Synchronization (GTDS):
	\be
	\frac{d{\bf X}}{dt}={\bf F}({\bf X})-\Dirac{\bf H}({\bf X})\, .
	\label{GD}
	\ee
where ${\bf F}({\bf X})$ and ${\bf H}({\bf X})$ have a block structure \be
{\bf F}({\bf X})=\begin{pmatrix}
			{{\bf f}({\bf x}^{(0)})}\\
			{{\bf f}({\bf x}^{(1)})}\\
			{{\bf f}({\bf x}^{(2)})}
		\end{pmatrix}\text{ and }\quad {\bf H}({\bf X})=\begin{pmatrix}
			{{\bf h}({\bf x}^{(0)})}\\
			{{\bf h}({\bf x}^{(1)})}\\
			{{\bf h}({\bf x}^{(2)})}
		\end{pmatrix}\, .
	\ee
Here, ${\bf f}({\bf x}^{(k)})$ and ${{\bf h}}({\bf x}^{(k)})$ are acting on each element of ${\bf x}^{(k)}$ as
	\be
	{\bf f}({\bf x}^{(k)})=\begin{pmatrix}
			{\vec{ f}(\vec{x}^{(k)}_1)}\\
			\vdots\\
			{\vec{f}(\vec{x}^{(k)}_i)}\\
			\vdots\\
			{\vec{f}(\vec{x}^{(k)}_{N_k})}\\
		\end{pmatrix}\text{ and } {\bf h}({\bf x}^{(k)})=\begin{pmatrix}
			{\vec{h}(\vec{x}^{(k)}_1)}\\
			\vdots\\
			{\vec{h}(\vec{x}^{(k)}_i)}\\
			\vdots\\
			{\vec{h}(\vec{x}^{(k)}_{N_k})}\\
		\end{pmatrix}\, ,
	\ee
 with $\vec{f}(\vec{x}^{(k)}_i)\in \mathbb{R}^d$, $\vec{h}(\vec{x}^{(k)}_i)\in \mathbb{R}^d$. Note that $\vec{f}$ and $\vec{h}$ can be any  arbitrary odd nonlinear functions. Indeed only odd $\vec{f}$ and $\vec{h}$ functions can  preserve the equivariance of the dynamics with respect to the choice of the orientation of the simplices, as discussed  in Ref.~\cite{carletti2023global}. Let us observe that, while the functions $\vec{f}$ should be the same for all $k$-dimensional simplex, $\sigma^k_i$, because we are considering identical topological dynamical systems, the functions $\vec{h}$ could depend on the simplex index $i$, denoting thus heterogeneity in the coupling. However for a sake of clarity, we prefer in the following to adopt the simplified assumption of homogeneous coupling, but the proposed results can be easily extended as to consider the more general framework. 
It follows that, in absence of the Dirac coupling, when $\bm\gamma_{[1]}=\bm\gamma_{[2]}=0$, the dynamics  for each topological signal $\vec{x}_i^{(k)}$ is identical 
\begin{equation}
\frac{d{\vec{x}}^{(k)}_i}{dt}=\vec{f}(\vec{x}^{(k)}_i)\, .
\label{isolated}
\end{equation}
 Namely, as already stated, the reaction term ${\vec{f}}$ determining the evolution of the each topological signal is independent of the other topological signals. In this way, the isolated system given by Eq.~\eqref{isolated}, and obtained when we silence the Dirac operator ${\Dirac}$, refers to each individual and isolated simplex dynamics.

When we take into consideration the action of the Dirac operator $\Dirac$, the explicit dynamics reads instead
\bea
\label{eq:coupleddirac}
\frac{d \vec{x}^{(0)}_i}{dt} = \vec{f}(\vec{x}^{(0)}_i)-\bm  \gamma_0^{(1)}\sum_{j=1}^{N_{1}} { B}_{1}(i,j) \vec{h}(\vec{x}^{(1)}_j),\nonumber\\
\frac{d \vec{x}^{(1)}_i}{dt} = \vec{f}(\vec{x}^{(1)}_i)-\bm  \gamma_1^{(2)}\sum_{j=1}^{N_{2}} { B}_{2}(i,j) \vec{h}(\vec{x}^{(2)}_j)-\bm  \gamma_1^{(1)}\sum_{j=1}^{N_{1}} { B}_{1}^\top(i,j) \vec{h}(\vec{x}^{(0)}_j),\nonumber \\
\frac{d \vec{x}^{(2)}_i}{dt} = \vec{f}(\vec{x}^{(2)}_i)-\bm  \gamma_2^{(2)}\sum_{j=1}^{N_{1}} {B}_{2}^\top(i,j) \vec{h}(\vec{x}^{(1)}_j).
\eea 
Note that this system can be generalized to $K>2$ simplicial and cell complexes in a straightforward way.

\subsection{Topological conditions for the existence of a  GTDS}
\label{sec:cond}
Let us indicate with $\vec{s}(t)$ a stable  solution of the autonomous system
 \be
 \frac{d \vec{s}}{dt} = \vec{f}(\vec{s})\, .
 \ee
 The Global Topological Dirac Synchronization (GTDS) is a state in which  the topological spinor is given by  \be{\bf X}=\bm \Phi=\hat{\bf U}\otimes \vec{s},\label{GTDSs}\ee where $\hat{\bf U}$ is a $\mathcal{N}$ column vector of elements $\hat{U}_i^{(0)}=1$ and $\hat{U}_i^{(k)}\in \{-1,1\}$ for $0<k\leq K$.
 Therefore, the GTDS state is characterized by the identical dynamics of each topological signal allowing only for a possible change of sign for topological signals of dimension $k>0$, whose sign depends on the cell orientation, i.e., \be\vec{x}_i^{(k)}=\hat{U}_i^{(k)}\vec{s},\quad  \forall i,\forall k\, .
 \ee
The dynamical system for topological signals coupled by the Dirac operator, Eq.~\eqref{GD}, admits a Global Topological Dirac 
 Synchronization state  if and only if  the Dirac operator $\Dirac$ admits in its kernel the topological spinor $\Phi$, i.e.,
	\be
	\Dirac {\bm\Phi}={\bf 0}\, .\label{condition_Dirac}\ee 
 Since $\vec{s}$ can be any arbitrary solution of the autonomous system, the latter equation implies that 
 \be
 {\bf D}\hat{\bf U}=0 \quad \Longrightarrow \quad  {\bm{\mathcal{L}}}\hat{\bf U}=0
 \label{conditions}
 \ee
	This condition limits the topologies that can sustain GTDS. This condition is specific to higher-order topological signals of dimension $k>0$ and does not have an equivalent for global node synchronization. An analogous condition is necessary for observing Global Topological Synchronization of $k$-dimensional topological signals with $k>0$ taken in isolation (see for details \cite{carletti2023global}).
 In the following paragraphs, we will discuss specifically the constraints that this condition imposes on $K=1$ dimensional simplicial complexes (networks) and on $K=2$ dimensional cell and simplicial complexes.
 

\subsection{Dynamical conditions for the existence of a  GTDS: the Master Stability Function (MSF) approach}

In this paragraph we investigate the stability of the GTDS state for the  dynamics defined in Eq.~\eqref{GD} under the hypothesis that such GTDS state exists.
In order to perform the stability analysis, we consider the Master Stability Function (MSF) approach~\cite{fujisaka1983stability,Pecora,Pecora_etal97,barahona2002synchronization}.
We thus expand the dynamical system Eq.~\eqref{GD} close to the GTDS state  by considering the ``perturbed'' topological spinor ${\bf X}=\bm\Phi+\delta{\bf X}$, by obtaining in this way the following linear dynamical system
	\be
	\frac{d \delta{\bf X}}{dt}=\left[{\boldsymbol{\mathcal{J}}_{\bf f}}- 
 \Dirac{\boldsymbol{{\mathcal{J}}_{\bf h}}}\right]\delta{\bf X}\, ,
	\label{deltaPsi}
	\ee
	where ${\boldsymbol{\mathcal{J}}}_{\bf f}$ and ${\boldsymbol{\mathcal{J}}}_{\bf h}$ are the matrices 
		\begin{align*}
  \label{Jf_Jh}
	{\boldsymbol{\mathcal{J}}}_{\bf f}=\begin{pmatrix}
			{\bf I}_{N_0}\otimes\mathbf{J}_{\vec{f}} & 0 & 0\\
			0 & {\bf I}_{N_1}\otimes\mathbf{J}_{\vec{f}} & 0\\
			0 & 0 & {\bf I}_{N_2}\otimes\mathbf{J}_{\vec{f}}
		\end{pmatrix}\, ,\nonumber \\ {\boldsymbol{\mathcal{J}}}_{\bf h}=\begin{pmatrix}
			{\bf I}_{N_0}\otimes\mathbf{J}_{\vec{h}} & 0 & 0\\
			0 & {\bf I}_{N_1}\otimes\mathbf{J}_{\vec{h}} & 0\\
			0 & 0 & {\bf I}_{N_2}\otimes\mathbf{J}_{\vec{h}}
		\end{pmatrix}\,,
	\end{align*}
	with $\mathbf{J}_{\vec{f}}$ and $\mathbf{J}_{\vec{h}}$ indicating the Jacobians of the functions ${\vec{f}}$ and ${\vec{h}}$, respectively, both computed on the topological signal $\vec{s}$, i.e., the GTDS state. 

	Resorting to the Dirac decomposition, i.e., Eq.~\eqref{Diracdec}, implying that $\delta {\bf X}=\delta {\bf X}_{[1]}+\delta {\bf X}_{[2]}+\delta {\bf X}^{\textrm{harm}}$, allows to greatly simplify the investigation of the stability of Eq.~\eqref{deltaPsi}. Indeed, for a $K=2$ dimensional simplicial and cell complex, the dynamical system Eq.~\eqref{deltaPsi} can be decomposed into two independent dynamical systems, one for $\delta{\bf X}_{[1]}$ and the other for $\delta {\bf X}_{[2]}$, that can be investigated independently.    
	Specifically, the dynamics for ${\delta{\bf X}}_{[k]}$ with $k\in \{1,2\}$ and ${\bf X}^{\textrm{harm}}$ can be obtained by starting from Eq. (\ref{deltaPsi}) and read (see  \ref{ApA} for details)
	\bea
	\hspace{15mm}\frac{d\delta{\bf X}_{[k]}}{dt}&=&\left[ {\boldsymbol{\mathcal{J}}}_{\bf f}^{[k]}-\bm\gamma_{[k]}{\boldsymbol{\mathcal{D}}}_{[k]}\otimes {\boldsymbol{{\mathcal{J}}}_{\bf h}}^{[k]}\right]\delta{\bf X}_{[k]}\, ,\label{decPsi} \\
	\hspace{15mm}\frac{d\delta{\bf X}^{\textrm{harm}}}{dt}&=&{\boldsymbol{\mathcal{J}}_{\bf f}}\delta{\bf X}^{\textrm{harm}}\, ,
	\label{31}
 \eea
 where we have indicated with  ${\boldsymbol{\mathcal{J}}}_{\bf f}^{[k]}$ and ${\boldsymbol{\mathcal{J}}}_{\bf h}^{[k]}$ the matrices
 \bea
	\hspace{-15mm}{\boldsymbol{\mathcal{J}}}_{\bf f}^{[1]}=\begin{pmatrix}
			{\bf I}_{{N_0}}\otimes\mathbf{J}_{\vec{f}} & 0 & 0\\
			0 & {\bf I}_{{N_1}}\otimes\mathbf{J}_{\vec{f}} & 0\\
			0 & 0 & 0
		\end{pmatrix}\, ,\quad {\boldsymbol{\mathcal{J}}}_{\bf h}^{[1]}=\begin{pmatrix}
			{\bf I}_{{N_0}}\otimes\mathbf{J}_{\vec{h}} & 0 & 0\\
			0 & {\bf I}_{{N_1}}\otimes\mathbf{J}_{\vec{h}} & 0\\
			0 & 0 & 0
		\end{pmatrix}\, ,\nonumber \\
		\hspace{-15mm}{\boldsymbol{\mathcal{J}}}_{\bf f}^{[2]}=\begin{pmatrix}
			0 & 0 & 0\\
			0 & {\bf I}_{{N_1}}\otimes\mathbf{J}_{\vec{f}} & 0\\
			0 & 0 & {\bf I}_{{N_2}}\otimes\mathbf{J}_{\vec{f}}
		\end{pmatrix}\, ,\quad {\boldsymbol{\mathcal{J}}}_{\bf h}^{[2]}=\begin{pmatrix}
			0 & 0 & 0\\
			0 & {\bf I}_{{N_1}}\otimes\mathbf{J}_{\vec{h}} & 0\\
			0 & 0 & {\bf I}_{{N_2}}\otimes\mathbf{J}_{\vec{h}}
		\end{pmatrix}\, .
	\eea
 The Global Topological Dirac synchronization will be stable (under small perturbations) if and only if the maximum Lyapunov exponent of the system in Eqs.(\ref{decPsi}) and { Eq. (\ref{31}) }is negative.
	Note that, by following a straightforward generalization of the argument, it is possible to study the GTDS on any arbitrary $K$ simplicial and cell complex by focusing on $K$ independent dynamical systems of the type of Eq.~\eqref{decPsi}-\eqref{31}.

\section{Dynamical system theory of GTDS on $K=1$ and $K=2$ dimensional simplicial complexes}

\subsection{Global Dirac Synchronization  on a 1-dimensional simplicial complex}
For the sake of concreteness, let us consider in detail the case of a $1$-simplicial complex, i.e., a simple network where nodes and edges signals interact via the Dirac operator $\Dirac=\bm\gamma_{[1]}{\bf D}_{[1]}$. Let $\vec{u}_i=\vec{x}^{(0)}_i\in\mathbb{R}^d$, resp. $\vec{v}_\ell=\vec{x}^{(1)}_\ell\in\mathbb{R}^d$, be the topological signals defined on the $i$-th node, resp. $\ell$-th edge, of a $1$-simplicial complex; then, the general system~\eqref{eq:coupleddirac} rewrites as
\begin{align}
 \frac{d\vec{u}_i}{dt} &= \vec{ f}(\vec{u}_i)-\bm \gamma_0^{(1)}\sum_{q=1}^{N_1}B_1(i,q)\vec{h}(\vec{v}_q)\\
  \frac{d\vec{v}_\ell}{dt} &= \vec{f}(\vec{v}_\ell)-\bm \gamma_1^{(1)}\sum_{j=1}^{N_0}B^\top_1(\ell,j)\vec{h}(\vec{u}_j)\, .
\label{eq:Diraccoup}
\end{align}
From this equation, we observe that the nodes evolution depends on the edge signals only through the Dirac operator and, vice-versa, the isolated system obtained when we silence the Dirac operator defines an independent dynamics of the topological signal defined on each node and each edge of the network.

 The GTDS state of this dynamics exists only if  the conditions Eq.~\eqref{conditions} are met.
 By considering here exclusively $K=1$ dimensional simplices, i.e., we impose ${\bf B}_2=0$, these conditions are met if and only if the network is Eulerian meaning that it has all the nodes of even degree, as observed in \cite{giambagli2022diffusion}.

The stability of the GTDS is determined by the MSF approach and, in general, by the systems provided by Eq.~\eqref{decPsi} and Eq.~\eqref{31}.
The stability of the harmonic signal ${\bf X}^{\textrm{harm}}$ is ensured by considering a stable solution $\vec{s}$ of the isolated dynamical system given by Eq.~\eqref{isolated}.
Thus, for a $1$-dimensional simplicial complex we need only to guarantee that Eq.~\eqref{decPsi} with $k=1$ has a negative Lyapunov exponent.
Thus, by defining $\delta\vec{u}_i(t)= \delta x^{(0)}_i(t)$ and $\delta\vec{v}_\ell(t)= \delta {x}_\ell^{(1)}(t)$, the linearized dynamics Eq.~\eqref{decPsi} can be rewritten, in the case under consideration $n=1$, as 
\begin{align}
\label{eq:Diraccouplin}
 \frac{d\delta\vec{u}_i}{dt} &= \mathbf{J}_{\vec{f}}\delta\vec{u}_i-\bm \gamma_0^{(1)}\sum_{q=1}^{N_1}B_1(i,q)\mathbf{J}_{\vec{h}}\delta\vec{v}_q\notag\\
  \frac{d\delta\vec{v}_\ell}{dt} &= \mathbf{J}_{\vec{f}}\delta\vec{v}_\ell-\bm \gamma_1^{(1)}\sum_{j=1}^{N_0}B^\top_1(\ell,j)\mathbf{J}_{\vec{h}}\delta\vec{u}_j\, ,
\end{align}
where $\mathbf{J}_{\vec{f}}$ is the Jacobian of ${\bf f}$ evaluated on the reference GTDS solution $\vec{s}(t)$.

Let $\psi_0^{{(\alpha)}}$, resp. $\psi_1^{{(\alpha)}}$, be the singular vectors of ${\bf B}_1$ corresponding to a non-zero singular value $b_{\alpha}$, thus satisfying 
\begin{equation}
\label{eq:relBpsi}
    \mathbf{B}_1\psi_1^{{(\alpha)}}=b_{\alpha}\psi_0^{{(\alpha)}}\, , \quad \mathbf{B}_1^{\top}\psi_0^{{(\alpha)}}=b_{\alpha}\psi_1^{{(\alpha)}}\, . 
\end{equation}  
It can be easily proved that $\psi_0^{{(\alpha)}}$, resp. $\psi_1^{{(\alpha)}}$, are eigenvectors of $\mathbf{L}_0$, resp. $\mathbf{L}_1$, associated to nonzero eigenvalues,  $\Lambda_0^{{(\alpha)}}=\Lambda_1^{{(\alpha)}}=b_{\alpha}^2$.
We notice that we can project the perturbations $\delta\vec{u}=(\delta\vec{u}_1,\dots,\delta\vec{u}_{N_0})^\top$ and $\delta\vec{v}=(\delta\vec{v}_1,\dots,\delta\vec{v}_{N_1})^\top$, as well as $\mathbf{B}_1^\top\delta\vec{u}$ and $\mathbf{B}_1\delta\vec{v}$, onto the basis of the eigenvectors $\psi_n^{{(\alpha)}}$, $n=0,1$, by defining thus $\delta\hat{u}_{\alpha}$ and $\delta \hat{v}_{\alpha}$ as follows
\be
\label{eq:projdeltas}
	\langle \psi_0^{(\alpha)},\delta \vec{u}\rangle=\delta \hat{u}_{\alpha} , \quad \langle \psi_1^{(\alpha)},\delta \vec{v}\rangle=\delta\hat{v}_{\alpha}\, . \ee
 where $\langle\cdot,\cdot\rangle$ denotes the scalar product. By using the latter equations, we obtain the following useful relations
 \be
		\langle \psi_0^{(\alpha)},\mathbf{B}_1\delta \vec{v}\rangle= b_{\alpha}\delta\hat{v}_{\alpha} , \quad \langle \psi_1^{(\alpha)},\mathbf{B}_1^{\top}\delta \vec{u}\rangle= b_{\alpha}\delta\hat{u}_{\alpha} \, .
\ee
Making use of the latter equations, we can rewrite Eq.~\eqref{eq:Diraccouplin} { }{ for each separate eigenmode $\alpha$} as follows:
\begin{align}
\label{eq:Diraccouplinproj}
 \frac{d\delta\hat{u}_{\alpha}}{dt} &= \mathbf{J}_{\vec{f}}\delta\hat{u}_{\alpha} - \bm \gamma_0^{(1)} b_{\alpha} \mathbf{J}_{\vec{h}}\delta\hat{v}_{\alpha}\notag \\
 \frac{d\delta\hat{v}_{\alpha}}{dt} &= \mathbf{J}_{\vec{f}}\delta\hat{v}_{\alpha} - \bm \gamma_1^{(1)} b_{\alpha} \mathbf{J}_{\vec{h}}\delta\hat{u}_{\alpha}\, ,
\end{align}
or, in matrix form by introducing $\delta_{\alpha} = (\delta\hat{u}_{\alpha},\delta\hat{v}_{\alpha})^\top$,
\begin{equation}
\label{eq:DiraccouplinprojMat}
\frac{d\delta_{\alpha}}{dt} = \left(
\begin{matrix}
 \mathbf{J}_{\vec{f}} & -b_{\alpha}  \bm \gamma_0^{(1)} \mathbf{J}_{\vec{h}}\\
 -b_{\alpha} \bm \gamma_1^{(1)} \mathbf{J}_{\vec{h}} & \mathbf{J}_{\vec{f}}
\end{matrix}
 \right)\delta_{\alpha}=:\mathbf{M}(b_{\alpha})\delta_{\alpha}\, .
 \end{equation} 
This equation allows to investigate the stability of the synchronous solution by studying the largest Lyapunov exponent of the above linear (non-autonomous) system. Let us observe that the matrix determining this linear system depends on the singular values of the boundary operator ${\bf B}_1$;  however, the following proposition allows us to prove that the spectrum of the matrix will depend only on $b_{\alpha}^2$, i.e., on the eigenvalues of the Laplacian.
\begin{proposition}
\label{prop:coeffb2}
Let us consider a square matrix of the form
\begin{equation}
 \mathbf{M}(w)= \left(
\begin{matrix}
 \mathbf{A}_1 & w   \mathbf{A}_2\\
 w \mathbf{A}_3 &  \mathbf{A}_4
\end{matrix}
 \right)
\end{equation}
where $ \mathbf{A}_i$, $i=1,2,3,4$, are four generic square matrices and $x$ a real parameter. Then for any integer $k$ we have
\begin{equation}
 \mathbf{M}^{2k}(w)= \left(
\begin{matrix}
 p^{(k)}_1(w^2) & w p^{(k)}_2(w^2)\\
 w p^{(k)}_3(w^2) &  p^{(k)}_4(w^2)
\end{matrix}
 \right)
\end{equation}
where $p_1^{(k)}(t)$ and $p_4^{(k)}(t)$, resp. $p_2^{(k)}(t)$ and $p_3^{(k)}(t)$, are polynomials of degree $k$, resp. $k-1$, in the variable $t$, with matrix coefficients depending on the matrices $\mathbf{A}_i$.
\end{proposition}

\textcolor{black}{The proof can be done by recurrence on the integer $k$ (see~\ref{sec:profprop})}. Moreover to determine the spectrum of $\mathbf{M}(b_{\alpha})$ given by~\eqref{eq:DiraccouplinprojMat}, we have to solve $\det (\mathbf{M}(b_{\alpha}) - \lambda\mathbf{I})=0$. The Cayley–Hamilton theorem allows to express the determinant of a matrix as linear combination of the trace of the powers of such matrix. By applying the previous Proposition to the matrix $\mathbf{M}(b_{\alpha}) - \lambda\mathbf{I}$, we can conclude that the trace of the powers of such matrix are polynomials in $b_{\alpha}^2$. Thus, the stability of the GTDS solution will depend on $b_{\alpha}^2=\Lambda^{(\alpha)}_0=\Lambda^{(\alpha)}_1>0$.

\subsection{ Stuart-Landau model on a 1-dimensional simplicial complex}
\label{sec:couplingDiracSL}

The aim of this section is to present the above analysis { by using as reference dynamical system the Stuart-Landau model, a paradigmatic model in the study of synchronization dynamics \cite{nakao2014complex}, which is representative of all oscillatory systems undergoing a supercritical Hopf-Andronov bifurcation~\cite{kuramoto2019concept,Kuramoto}.} More precisely, we assume to have a complex topological signal defined on each node, $u_j(t)\in \mathbb{C}$, and a second complex topological signal defined on edges, $v_\ell(t)\in \mathbb{C}$, whose evolution is described by
\begin{align}
\label{eq:DiracSL}   
\frac{d {u}_j}{dt} &= \sigma u_j-\beta u_j|u_j|^2-\mu^{(0)}\sum_{q=1}^{N_1}B_1(j,q)v_q\\
\frac{d {v}_\ell}{dt} &= \sigma v_\ell-\beta v_\ell|v_\ell|^2-\mu^{(1)}\sum_{j=1}^{N_0}B_1^\top(\ell,j)u_j \, ,
\end{align}
where $\sigma$ and $\beta$ are the complex Stuart-Landau parameters, and $\mu^{(a)}=\gamma_a^{(1)}\in \mathbb{C}$, $a=0,1$, are the complex coupling strengths for nodes and links that play the role of the gamma matrices. Let us observe that we assumed linear coupling functions, i.e., $\vec{ h}(\vec{x})=\vec{x}$, nonetheless, the following analysis holds true in a more general setting.

The system defined on nodes and on edges admits a limit cycle solution 
{ 
\begin{align}\hat{z}(t)=\sqrt{\frac{\sigma_\Re}{\beta_\Re}}e^{i\omega t},\end{align} }
where $\omega=\sigma_\Im-\beta_\Im \sigma_\Re/\beta_\Re$. Such a solution is stable if $\sigma_\Re>0$ and $\beta_\Re>0$, conditions that we hereby assume to hold true. To study the emergence of global topological Dirac synchronization, we perturb the above limit cycle solution and we study the time evolution of the perturbation: more precisely, we set 
{ 
\begin{align}u_j=\hat{z}(1+\rho_j)e^{i\theta_j},\quad \mbox{for}\  j=1,\dots, N_0, \nonumber \\ v_\ell=\hat{z}(1+\eta_\ell)e^{i\varphi_\ell},\quad \mbox{for}\  \ell=1,\dots, N_1,\end{align}} 
where $\rho_j$, $\theta_j$, $\eta_\ell$ and $\varphi_\ell$ are ``small'' real functions. 
 After some straightforward computations, we eventually obtain
\begin{align}
\label{eq:maineqDiracSLlin}
\frac{d\rho_j}{dt} &= -2\sigma_\Re \rho_j  -\sum_{q=1}^{N_1} B_1(j,q)\left(\mu^{(0)}_\Re \eta_q-\mu^{(0)}_\Im \varphi_q\right)\nonumber \\
\frac{d\theta_j}{dt} &= -2\beta_\Im\frac{\sigma_\Re}{\beta_{\Re}} \rho_j  -\sum_{q=1}^{N_1} B_1(j,q)\left(\mu^{(0)}_\Im \eta_q+\mu^{(0)}_\Re \varphi_q\right)\nonumber\\
\frac{d\eta_\ell}{dt} &= -2\sigma_\Re \eta_\ell  -\sum_{j=1}^{N_0} B_1^\top(\ell,j)\left(\mu^{(1)}_\Re \rho_j-\mu^{(1)}_\Im \theta_j\right)\nonumber\\
\frac{d\varphi_\ell}{dt} &= -2\beta_\Im\frac{\sigma_\Re}{\beta_{\Re}} \eta_\ell  -\sum_{j=1}^{N_0} B^\top_1(\ell,j)\left(\mu^{(1)}_\Im \rho_j+\mu^{(1)}_\Re \theta_j\right)\, .
\end{align}
Let us  project the vectors $\vec{\rho}=(\rho_1,\dots,\rho_{N_0})^\top$ and $\vec{\theta}=(\theta_1,\dots,\theta_{N_0})^\top$  on the singular vector $\psi_0^{(\alpha)}$,  and the vectors $\vec{\eta}=(\eta_1,\dots,\eta_{N_1})^\top$ and $\vec{\varphi}=(\varphi_1,\dots,\varphi_{N_1})^\top$ onto the singular vector $\psi_1^{(\alpha)}$, by defining in this way the quantities 
\begin{align}
\label{eq:projdeltas2}
	\langle \psi_0^{(\alpha)}, \vec{\rho}\rangle= \hat{\rho}_{\alpha} , \quad \langle \psi_0^{(\alpha)},\vec{\theta}\rangle=\hat{\theta}_{\alpha}, \\
 \langle \psi_1^{(\alpha)}, \vec{\eta}\rangle= \hat{\eta}_{\alpha} , \quad \langle \psi_1^{(\alpha)}, \vec{\varphi}\rangle=\hat{\varphi}_{\alpha}. 
 \end{align}
In terms of these new variables, the dynamical system in Eq.~\eqref{eq:maineqDiracSLlin} can be expressed as 
\begin{align}
\label{eq:maineqDiracSLlinproj}
\frac{d\hat{\rho}_\alpha}{dt} &= -2\sigma_\Re \hat{\rho}_\alpha -b_\alpha\left(\mu^{(0)}_\Re \hat{\eta}_\alpha-\mu^{(0)}_\Im \hat{\varphi}_\alpha\right)\nonumber\\
\frac{d\hat{\theta}_\alpha}{dt} &= -2\beta_\Im\frac{\sigma_\Re}{\beta_{\Re}} \hat{\rho}_\alpha  -b_\alpha \left(\mu^{(0)}_\Im \hat{\eta}_\alpha+\mu^{(0)}_\Re \hat{\varphi}_\alpha\right)\nonumber\\
\frac{d\hat{\eta}_\alpha}{dt} &= -2\sigma_\Re \hat{\eta}_\alpha  -b_\alpha\left(\mu^{(1)}_\Re \hat{\rho}_\alpha-\mu^{(1)}_\Im \hat{\theta}_\alpha\right)\nonumber\\
\frac{d\hat{\varphi}_\alpha}{dt} &= -2\beta_\Im\frac{\sigma_\Re}{\beta_{\Re}} \hat{\eta}_\alpha  -b_\alpha\left(\mu^{(1)}_\Im \hat{\rho}_\alpha+\mu^{(1)}_\Re \hat{\theta}_\alpha\right)\, .
\end{align}
We can define $\mathbf{J}_\alpha = \left(
\begin{smallmatrix}
 \mathbf{J} &  b_\alpha \mathbf{M}^{(0)}\\
b_\alpha \mathbf{M}^{(1)} &  \mathbf{J}
\end{smallmatrix}\right)$
where $\mathbf{J} =\left(
\begin{smallmatrix}
 -2\sigma_\Re & 0\\
 -2\beta_\Im\frac{\sigma_\Re}{\beta_\Re}& 0
\end{smallmatrix}
\right)$ is the Jacobian matrix of the Stuart-Landau system evaluated on the limit cycle solution, and 
$\mathbf{M}^{(a)}=\left(
\begin{smallmatrix}
 -\mu^{(a)}_\Re & \mu^{(a)}_\Im\\
 -\mu^{(a)}_\Im & -\mu^{(a)}_\Re
\end{smallmatrix}
\right)$, for $a=0,1$, and thus rewrite Eq.~\eqref{eq:maineqDiracSLlinproj} as follows
\begin{equation}
 \frac{d}{dt}\left(
\begin{smallmatrix}
{\hat{\rho}}_\alpha\\{\hat{\theta}}_\alpha\\{\hat{\eta}}_\alpha\\{\hat{\varphi}}_\alpha
\end{smallmatrix}\right) = \mathbf{J}_\alpha\left( \begin{smallmatrix}
{\hat{\rho}}_\alpha\\{\hat{\theta}}_\alpha\\{\hat{\eta}}_\alpha\\{\hat{\varphi}}_\alpha
\end{smallmatrix}\right)\, .
 \nonumber \end{equation}
The spectrum of $\mathbf{J}_\alpha$ determines the (local) stability property of the solutions of Eq.~\eqref{eq:maineqDiracSLlinproj}. In particular, Global Topological Dirac Synchronization will emerge if $\lambda=\max_{\alpha,j} \Re \lambda_j(b_\alpha)$ is negative, where $\lambda_j(b_\alpha)$, $j=1,\dots,4$, are the four eigenvalues of $\mathbf{J}_\alpha$. The characteristic polynomial is given by
\begin{equation}
\label{eq:polcar}
p_\alpha(\lambda) = \det \left(\mathbf{J}_\alpha-\lambda \mathbf{I}_4\right)=a_0\lambda^4+a_1 \lambda^3+a_2 \lambda^2+a_3\lambda + a_4\, ,
\end{equation}
where the coefficients $a_j$, $j=0,\dots,4$ can be computed by using the Matlab~\cite{MATLAB} symbolic engine:
\begin{eqnarray}\label{eq:coeff}
a_0 &=& 1,\nonumber \\
a_1 &=& -2\mathrm{tr}\left(\mathbf{J}\right) = 4\sigma_\Re,\nonumber \\
a_2 &=& - 2b_\alpha^2 (\mu^{(0)} \mu^{(1)})_\Re + 4 (\sigma_\Re)^2, \nonumber\\
a_3 &=& -4 b_{\alpha}^2 \frac{\sigma_\Re}{\beta_\Re} \left(\beta_\Re (\mu^{(0)} \mu^{(1)})_\Re + \beta_\Im (\mu^{(0)} \mu^{(1)})_\Im\right), \\
a_4 &=& b_\alpha^4|\mu^{(0)}_\Re|^2 |\mu^{(1)}|^2  -4b_\alpha^2 \frac{\sigma_\Re^2}{\beta_\Re^2}\left(\beta_\Re^2\mu^{(0)}_\Re\mu^{(1)}_\Re +\beta_\Im\beta_\Re\mu^{(0)}_\Re\mu^{(1)}_\Im \right. \nonumber \\
    && \left. +\beta_\Im\beta_\Re\mu^{(0)}_\Im\mu^{(1)}_\Re +\beta_\Im^2\mu^{(0)}_\Im\mu^{(1)}_\Im\right).\nonumber
\end{eqnarray}
From the latter relations it follows that the coefficients $a_k$ depend on $b_\alpha^2$, as claimed in Proposition~\ref{prop:coeffb2}, hence the Master Stability Function for Dirac synchronization depends on the eigenvalues of the Laplace matrix as in the case of synchronization of node signals interacting with a diffusive coupling.

To study the stability of the polynomial $p_\alpha(\lambda)$, one could resort to the Routh-Hurwitz \cite{Routh1877,Hurwitz1895} criterion, as we show in~\ref{sec:ApC}. However, such method gives us the conditions under which the Master Stability Function (MSF) is always negative, which is a sufficient condition to obtain GTDS, but not necessary, due to discrete nature of the support. In fact, as we will show in the following, the MSF we are considering often assume both positive and negative signs, hence GTDS can emerge by a suitable choice of the simplicial complex, with the ``right'' singular values $b_\alpha$. 
 \begin{figure}[h]
\centering
\includegraphics[width=0.95\linewidth]{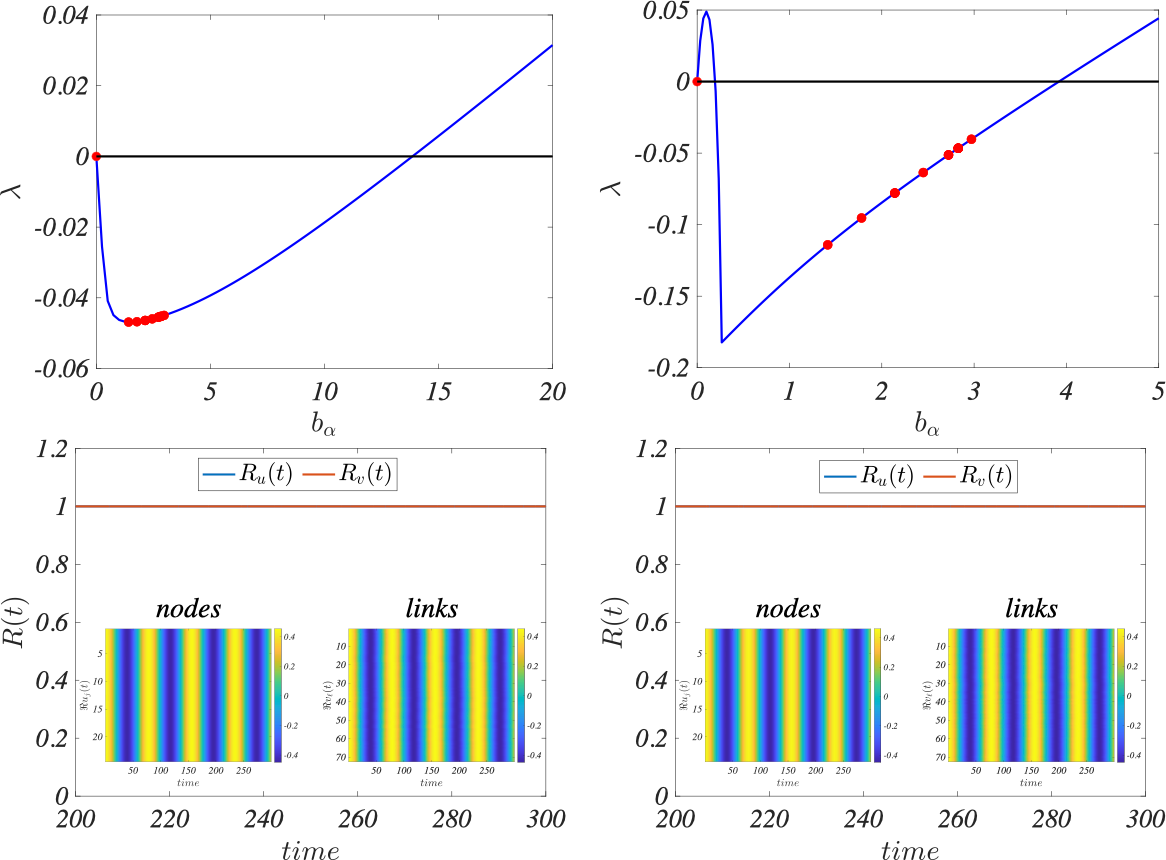}
\caption{Top panels: we report the Master Stability Function, i.e., the maximum of the real part of the characteristic roots, as a function of $b_\alpha$, the singular values of the matrix $\mathbf{B}_1$ ($C_1<0$ left column and $C_1>0$ right column, in both cases $\mu^{(0)}_\Im \mu^{(1)}_\Im>0$). Bottom panels show the order parameter for node and   edge signals together with the real part of the node and   edge signal, defined on a triangulated $2$-torus. In the left column panels we used the parameters $\mu^{(0)}_\Im = -0.5$, $\mu^{(1)}_\Re = -0.5$ and $\mu^{(1)}_\Im = -0.24$, while in the right column panels $\mu^{(0)}_\Im = -1.5$, $\mu^{(1)}_\Re = -0.75$ $\mu^{(1)}_\Im = -1.0$.  The remaining parameters are:
$\sigma_\Re = 0.2$, $\sigma_\Im = 0.3$, $\beta_\Re = 1.0$, $\beta_\Im = 1.1$, $\mu^{(0)}_\Re = 1.0$.}
\label{fig:numres}
\end{figure}

An alternative approach is based on the observation that the four roots of the characteristic polynomial~\eqref{eq:polcar} for $\alpha=1$, i.e., once we substitute $b_1=0$, are given by
\begin{equation}
 \lambda_1(b_1)= \lambda_2(b_1)=-2\sigma_\Re\quad \text{ and }\quad \lambda_3(b_1)= \lambda_4(b_1)=0\, .
\nonumber \end{equation}
This is because the reference solution is a stable limit cycle. For small $b_\alpha$ the roots $\lambda_1(b_\alpha)$ and $\lambda_2(b_\alpha)$ will (generically) assume different values but they remain negative. On the other hand, the vanishing roots, $\lambda_3(b_\alpha)$ and $\lambda_4(b_\alpha)$, will bifurcate from $0$ either by reaching positive real parts or negative ones; remember that those roots should be complex conjugate, the coefficients of the characteristic polynomial being real numbers. We can thus look for an expansion of $\lambda_3(b_\alpha)$ and $\lambda_4(b_\alpha)$ of the form $\lambda_j(b_\alpha)=\lambda_j^{(1)} b_\alpha+\lambda_j^{(2)} b_\alpha^2+\dots$, for $j=3,4$ and for small $b_\alpha$. A straightforward computation \textcolor{black}{(see~\ref{sec:perturbcalc})} returns
\begin{eqnarray}
\label{eq:rootapprox}
\lambda_3(b_\alpha)=  -\frac{\sqrt{C_1}}{\beta_\Re}b_\alpha-\frac{\mu^{(0)}_\Im \mu^{(1)}_\Im}{2\beta_\Re^2\sigma_\Re}|\beta|^2b_\alpha^2+\mathcal{O}(b_\alpha^3) \\
\lambda_4(b_\alpha)=  \frac{\sqrt{C_1}}{\beta_\Re}b_\alpha-\frac{\mu^{(0)}_\Im \mu^{(1)}_\Im}{2\beta_\Re^2\sigma_\Re}|\beta|^2b_\alpha^2+\mathcal{O}(b_\alpha^3)\, ,
\end{eqnarray}
where
\begin{equation}
C_1= (\beta_\Re \mu^{(0)}_\Re+\beta_\Im \mu^{(0)}_\Im)(\beta_\Re \mu^{(1)}_\Re+\beta_\Im \mu^{(1)}_\Im),
\nonumber \end{equation}
and we can draw the following conclusions. If $C_1<0$ then the roots are complex conjugate and moreover
\begin{equation}
 \Re(\lambda_j(b_\alpha)) = -\frac{\mu^{(0)}_\Im \mu^{(1)}_\Im}{2\beta_\Re^2\sigma_\Re}|\beta|^2b_\alpha^2+\mathcal{O}(b_\alpha^3)\, ,
 \nonumber
\end{equation}
hence $\Re(\lambda_j(b_\alpha))<0$ if $\mu^{(0)}_\Im \mu^{(1)}_\Im>0$. On the other hand, if $C_1>0$ the roots are real and, moreover, $\lambda_3(b_\alpha)<0$ and $\lambda_4(b_\alpha)>0$.

We can thus conclude that, if $C_1<0$ and $\mu^{(0)}_\Im \mu^{(1)}_\Im>0$, then $\lambda<0$ (see top left panel of Fig.~\ref{fig:numres}) and thus any simplicial complex for which $\max_\alpha b_\alpha$ is sufficiently small, will support the global synchronization of the Stuart-Landau system. This claim can be appreciated by looking at bottom left panel of Fig.~\ref{fig:numres}, where we report the order parameter for the topological node signal, $R_u(t)=\frac{1}{N_0}\lvert \sum_j u_j(t)/\hat{z}\rvert$, and  edge signal, $R_v(t)=\frac{1}{N_1}\lvert \sum_\ell v_\ell(t)/\hat{z}\rvert$, together with the time evolution of the real part of the nodes signal, $\Re u_j(t)$ (left inset), and the real part of the edges signal, $\Re v_\ell(t)$ (right inset), in the case of a $1$-simplicial complex obtained by realizing a triangular mesh on a $2$-torus~\footnote{One can prove that the $1$-simplicial complex obtained by triangulating a $2$-torus satisfies the conditions $\mathbf{B_1}(1,\dots,1)^\top=0$ and $\mathbf{B_1}^\top(1,\dots,1)^\top=0$, guaranteeing the existence of the GTDS.}.

If the condition on $C_1$ is violated, it can happen that $\lambda$ reaches negative values and then increases again (see top right panel of Fig.~\ref{fig:numres}). In this case the Stuart-Landau system defined on the $1$-simplicial complex can exhibit GTDS if $b_\alpha$ belongs to a suitable interval, as we can appreciate from the bottom right panel of Fig.~\ref{fig:numres}, where we report again the order parameter for nodes and edges signals together with the real part of the node and  edge signals, again in the case of a triangulated $2$-torus.
\begin{figure}[h]
\centering
\includegraphics[width=0.75\linewidth]{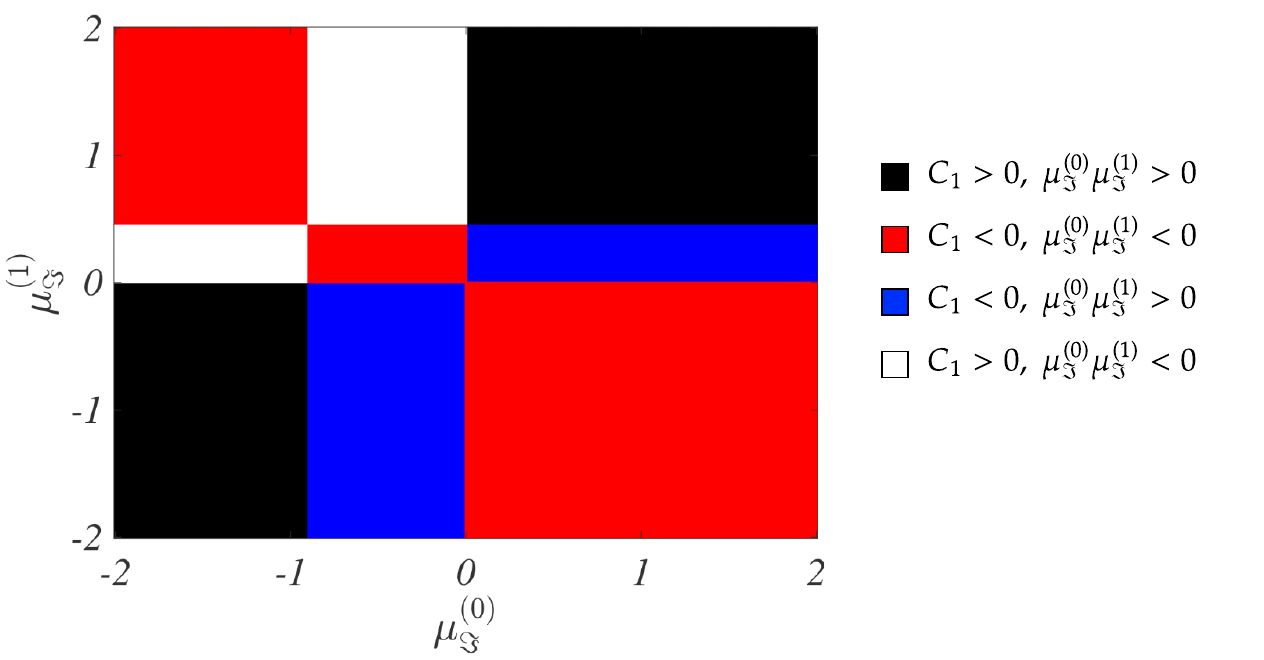}
\caption{We present the sign of the quantities $C_1$ and $\mu^{(0)}_\Im \mu^{(1)}_\Im$ as a function of $\mu^{(0)}_\Im$ and $\mu^{(1)}_\Im$. The remaining parameters have been fixed to $\sigma_\Re = 0.2$, $\sigma_\Im = 0.3$, $\beta_\Re = 1.0$, $\beta_\Im = 1.1$, $\mu^{(0)}_\Re = 1.0$, $\mu^{(1)}_\Re = -0.5$, namely the same used in the left columns of Fig.~\ref{fig:numres}. The red region corresponds to $C_1<0$ and $\mu^{(0)}_\Im \mu^{(1)}_\Im<0$, the white one to $C_1>0$ and $\mu^{(0)}_\Im \mu^{(1)}_\Im<0$, the blue one to $C_1<0$ and $\mu^{(0)}_\Im \mu^{(1)}_\Im>0$ and the black one to $C_1>0$ and $\mu^{(0)}_\Im \mu^{(1)}_\Im>0$. Let us observe that parameters associated to the blue region allow for global topological synchronization provided $\max b_\alpha$ is small enough.}
\label{fig:numresC1C2}
\end{figure}
Let us observe that, by fixing all the parameters but $\mu^{(0)}_\Im$ and $\mu^{(1)}_\Im$, the condition $C_1>0$ determines a ``chessboard''-like region with four parts, and the same holds true for the condition $\mu^{(0)}_\Im \mu^{(1)}_\Im>0$. In conclusion, the plane $(\mu^{(0)}_\Im, \mu^{(1)}_\Im)$ is divided into rectangular zones each one associated with a given sign of the above conditions. In Fig.~\ref{fig:numresC1C2}, we report the results for the parameters setting used in the left columns of Fig.~\ref{fig:numres}, i.e., $\sigma_\Re = 0.2$, $\sigma_\Im = 0.3$, $\beta_\Re = 1.0$, $\beta_\Im = 1.1$, $\mu^{(0)}_\Re = 1.0$, $\mu^{(1)}_\Re = -0.5$; by varying $\mu^{(0)}_\Im$ and $\mu^{(1)}_\Im$ in the range $[-2,2]$ we show the values of $C_1$ and $\mu^{(0)}_\Im \mu^{(1)}_\Im$ by using the following color code: $C_1<0$ and $\mu^{(0)}_\Im \mu^{(1)}_\Im<0$ red, $C_1>0$ and $\mu^{(0)}_\Im \mu^{(1)}_\Im<0$ white, $C_1<0$ and $\mu^{(0)}_\Im \mu^{(1)}_\Im>0$ blue and $C_1>0$ and $\mu^{(0)}_\Im \mu^{(1)}_\Im>0$ black. Hence, the blue and the black regions are associated to parameters values for which we can find $1$-simplicial complexes with suitable spectrum to ensure the emergence of global synchronization for topological Stuart-Landau defined on nodes and edges.

\begin{figure}
    \centering
    \includegraphics[width=0.7\linewidth]{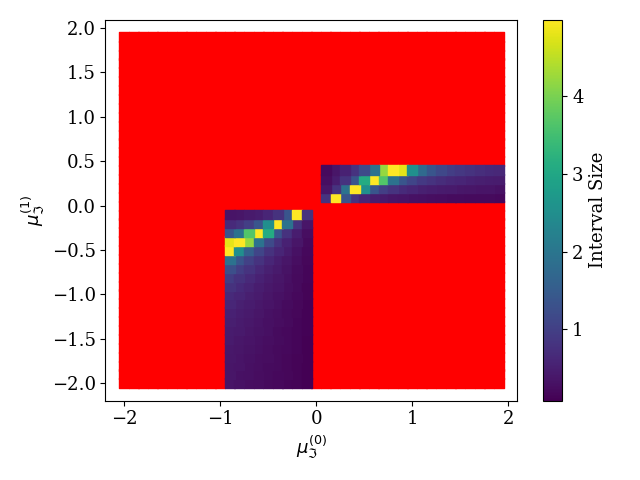}
    \caption{We present an extended analysis of the results reported in Fig.~\ref{fig:numresC1C2}, where the size of the interval of stability, $[0,\hat{b}]$, is depicted by using a color code. Red areas indicate values of $\mu_\Im^{(0,1)}$ that correspond to an unstable region near $0$, namely a positive dispersion relation. Blue values are associated to small $\hat{b}$, while yellow ones to large $\hat{b}$. Those results have been numerically obtained by using a grid search (data and more details are available in the paper repository \cite{github_repo}).}
    \label{fig:numres_StabilityRegion}
\end{figure}

Let us observe that from the these results we cannot assess the size of the interval, $[0,\hat{b}]$, for which the MSF is negative, the latter can however be numerically studied with hyperparameters optimization techniques, inherited from machine learning scenarios \cite{liaw2018tune} and shows a non-trivial behavior (see  Fig.~\ref{fig:numres_StabilityRegion}). In the latter figure we fix all the parameters but $\mu^{(0)}_\Im$ and $\mu^{(1)}_\Im$, we compute the MSF and we obtain the largest $\hat{b}$ for which the latter is negative. The values of $\hat{b}$ are reported by using a color code: blue means small $\hat{b}$ and thus the interval for which the system exhibits GTDS is relatively small, yellow stands for large values of $\hat{b}$ and thus a larger range of singular values for which the system shows GTDS. The red region corresponds to parameters for which the dispersion relation is positive (close to $0$) and thus GTDS can not be obtained except if the the singular values $b_\alpha$ are large enough.
 In Figure \ref{fig:numres_StabilityRegion}, the search for the stability region has been conducted over a $2$-dimensional domain by using a grid search; however, more advanced tools can be employed if a larger set of parameters has to be considered. The repository \cite{github_repo} allows for an easy implementation of such an extension.

\subsection{Dynamics of topological signal defined on weighted $2$-simplicial complexes}
\label{sec:w2sc}

In the previous sections we considered topological signals defined on nodes and edges of a $1$-simplicial complex, it is thus natural to study the dynamics of similar quantities in the case of a $2$-dimensional cell complex where, i.e., besides nodes and edges there are also polygons (triangles, squares, etc.). Let us thus consider the topological signals $\vec{u}_i=\vec{x}^{(0)}_i\in\mathbb{R}^d$,  $\vec{v}_\ell=\vec{x}^{(1)}_\ell\in\mathbb{R}^d$, and $\vec{z}_r=\vec{x}^{(2)}_r\in\mathbb{R}^d$ defined on the $i$-th node, the $\ell$-th edge, and the $r$-th polygon of a $2$-dimensional cell complex giving rise to the Global Topological Dirac Dynamics~\eqref{eq:coupleddirac} that we rewrite here for convenience,
 \begin{align}
 \frac{d\vec{u}_i}{dt} &= \vec{ f}(\vec{u}_i)-\bm\gamma_0^{(1)}\sum_{q=1}^{N_1}B_1(i,q)\vec{ h}(\vec{v}_q)\notag\\
  \frac{d\vec{v}_\ell}{dt} &= \vec{ f}(\vec{v}_\ell)-\bm\gamma^{(1)}_1\sum_{j=1}^{N_0}B^\top_1(\ell,j)\vec{h}^{(1)}_1(\vec{u}_j)-\bm\gamma^{(2)}_1\sum_{s=1}^{N_2}B_2(\ell,s)\vec{h}^{(2)}_1(\vec{z}_s)\label{eq:Diraccoup2}\\
    \frac{d\vec{z}_r}{dt} &= \vec{ f}(\vec{z}_r)-\bm\gamma_2^{(2)}\sum_{q=1}^{N_1}B^\top_2(r,q)\vec{ h}(\vec{v}_q)\notag\, ,
 \end{align}  
where we recall that the $2$-cell complex has $N_0$ nodes, $N_1$ links and $N_2$ polygons supporting topological signals.

As expressed in general terms in Sec.~\ref{sec:cond} on simplicial and cell complexes the existence of the GTDS is not guaranteed and only some specific higher-order network topologies admit this very homogeneous state. By assuming that the GTDS exists, its stability is dictated by the linearized systems Eq.~\eqref{decPsi} and Eq.~\eqref{31}.
The stability of the harmonic mode ${\bf X}^{\textrm{harm}}$ of the topological spinor is ensured by considering a stable solution $\vec{s}$ of the isolated system Eq.~\eqref{isolated}, condition that we assume to hold true. Thus the study of the stability of the GTDS reduces to the study of two independent dynamical systems, given by Eq.~\eqref{decPsi} obtained for $k=1$ and $k=2$ and defined on nodes and edges for $k=1$ and on edges and triangles for $k=2$.
These two linearized systems are both required to have a negative maximal Lyapunov exponent for ensuring the stability of the GTDS on the $K=2$ cell complex; let us observe that they can be studied independently of each other by using the same dynamical system theory discussed in the previous two paragraphs for the $K=1$ dimensional simplicial complex, namely to project on suitable basis of the involved subspaces.

\begin{figure}[h]
\centering
\includegraphics[width=\linewidth]{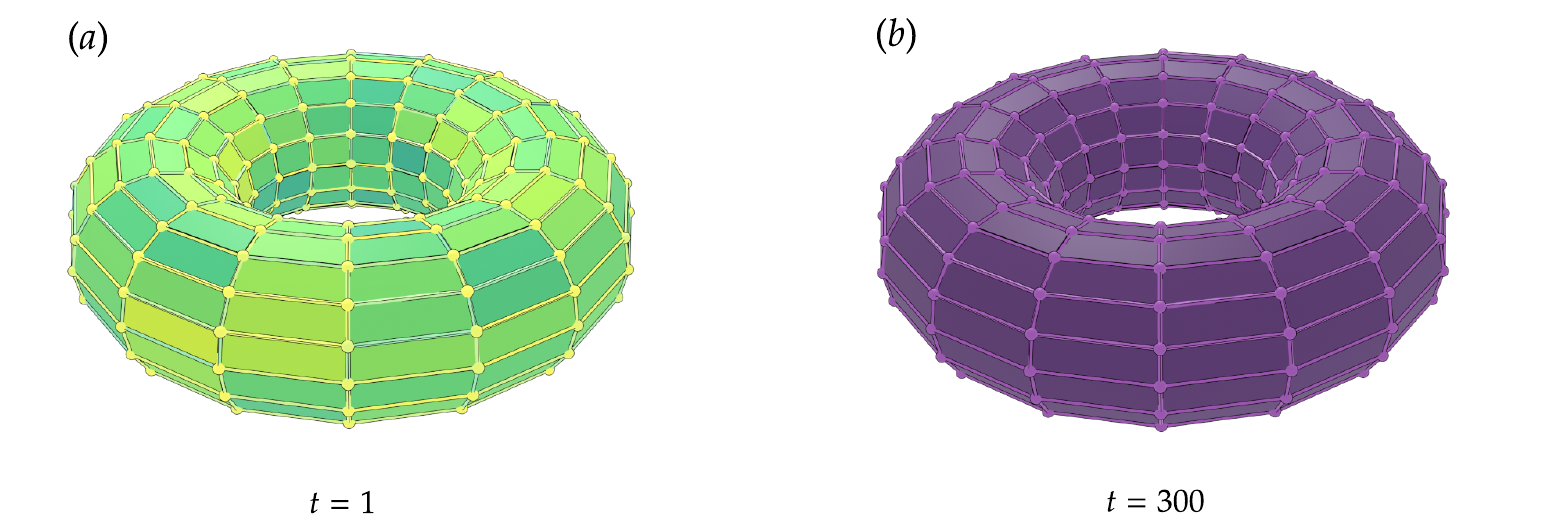}
\caption{Global Topological Dirac Synchronization (GTDS) is a viable dynamical state on a  Square Lattice Tessellation of the  $2$-Torus (SLTT) and occurs when all nodes, edges and squares signals follow the same dynamics. The latter is here represented during  (panel (a)) and after (panel (b)) the transient. The dynamics is taken to be driven by the Stuart Landau model. The chosen parameters ensure that the GTDS is stable and are here taken to be: $\sigma_\Re = 0.2$, $\sigma_\Im = 0.3$, $\beta_\Re = 1.0$, $\beta_\Im = 1.1$, $\mu^{(0)}_\Re = 1.0$, $\mu^{(0)}_\Im = -0.5$, $(\mu_{1}^{(1)})_\Re=(\mu_{1}^{(2)})_\Re=-0.5$, $(\mu_{1}^{(1)})_\Im=(\mu_{1}^{(2)})_\Im=-0.24$, $\mu^{(2)}_\Re = 1.0$, $\mu^{(2)}_\Im = -0.5$. See supplementary movie S1 to appreciate the temporal evolution of the topological signals on nodes, links and faces, toward global synchronisation.}
\label{fig:fig_dyn}
\end{figure}

An important question that arises in the study of GTDS on higher-order networks of dimension $K>1$ is whether this interesting dynamical state can be ever realized.
The topological conditions expressed in Eq.~\eqref{conditions} that a $K=2$ dimensional cell complex needs to satisfy to sustain GTDS are very stringent. However following similar arguments of Ref. \cite{carletti2023global} we can prove that the Square Lattice Tessellations of a $2$-dimensional Torus (SLTT)  admits this state and when GTDS is stable nodes, edges and square follows the same dynamics (see Figure \ref{fig:fig_dyn} for a visualization of GTDS). Interestingly, as a matter of fact, their arbitrary $K$-dimensional generalization also admit GTDS by involving all topological signals of dimension $0\leq k\leq K$. The interested reader can find a detailed derivation of the spectral properties of this $2$-cell complex in~\ref{ApA}.

\begin{figure}[h]
\centering
\includegraphics[width=0.8\linewidth]{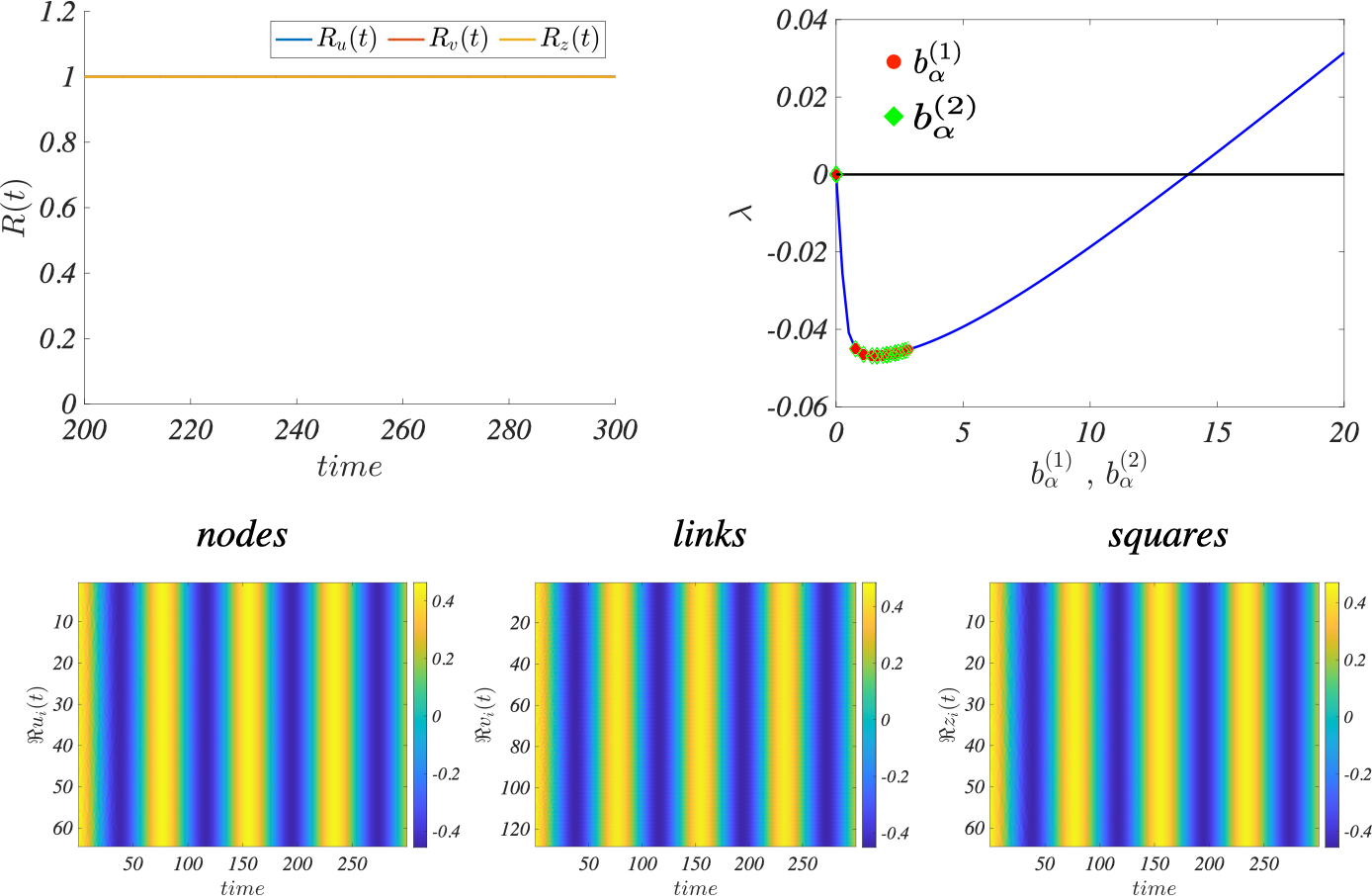}
\caption{Square tessellation of a $2$-Torus supporting Global Topological Synchronization. Top left panel: we show the order parameters for nodes, links and squares as a function of time.
Top right panel: we report the Master Stability Function, i.e., the maximum of the real part of the characteristic roots, as a function of $b_{\alpha}^{(1)}$ and $b_{\alpha}^{(2)}$, the singular values of the matrices $\mathbf{B}_1$ and $\mathbf{B}_2$. Bottom panels show the time evolution of (the real part of) the topological signal for nodes, links and squares. The used parameters are: $\sigma_\Re = 0.2$, $\sigma_\Im = 0.3$, $\beta_\Re = 1.0$, $\beta_\Im = 1.1$, $\mu^{(0)}_\Re = 1.0$, $\mu^{(0)}_\Im = -0.5$, $(\mu_{1}^{(1)})_\Re=(\mu_{1}^{(2)})_\Re=-0.5$, $(\mu_{1}^{(1)})_\Im=(\mu_{1}^{(2)})_\Im=-0.24$, $\mu^{(2)}_\Re = 1.0$, $\mu^{(2)}_\Im = -0.5$.}
\label{fig:SW2TorusSynch}
\end{figure}

We have considered the Stuart-Landau model on the 2-dimensional cell and simplicial complexes. The results reported in Figure~\ref{fig:SW2TorusSynch}  demonstrate that GTDS can emerge for topological signals defined on nodes, edges and squares of a square lattice tessellation of the $2$-torus under suitable dynamical conditions on the model parameters. 

We are, however, facing an issue once looking for GTDS on simplicial complexes of dimension $K>1$. Indeed, the conditions~\eqref{conditions} for the existence of a GTDS can never be satisfied on simplicial complexes of dimension $K>1$ as long as the simplicial complex is unweighted. Indeed a necessary condition for GTDS to occur is that edge signals will need to globally synchronize on a $K>1$ simplicial complex, and this has been shown to be impossible in Ref.~\cite{carletti2023global}.
Considering weighted simplicial complexes can however allow GTDS also on simplicial complexes of dimension $K>1$ similarly to what happens to global synchronization of topological signals of dimension $k$, as discussed in Ref.~\cite{WMCB2024}.

\begin{figure}[h]
\centering
\includegraphics[width=0.98\linewidth]{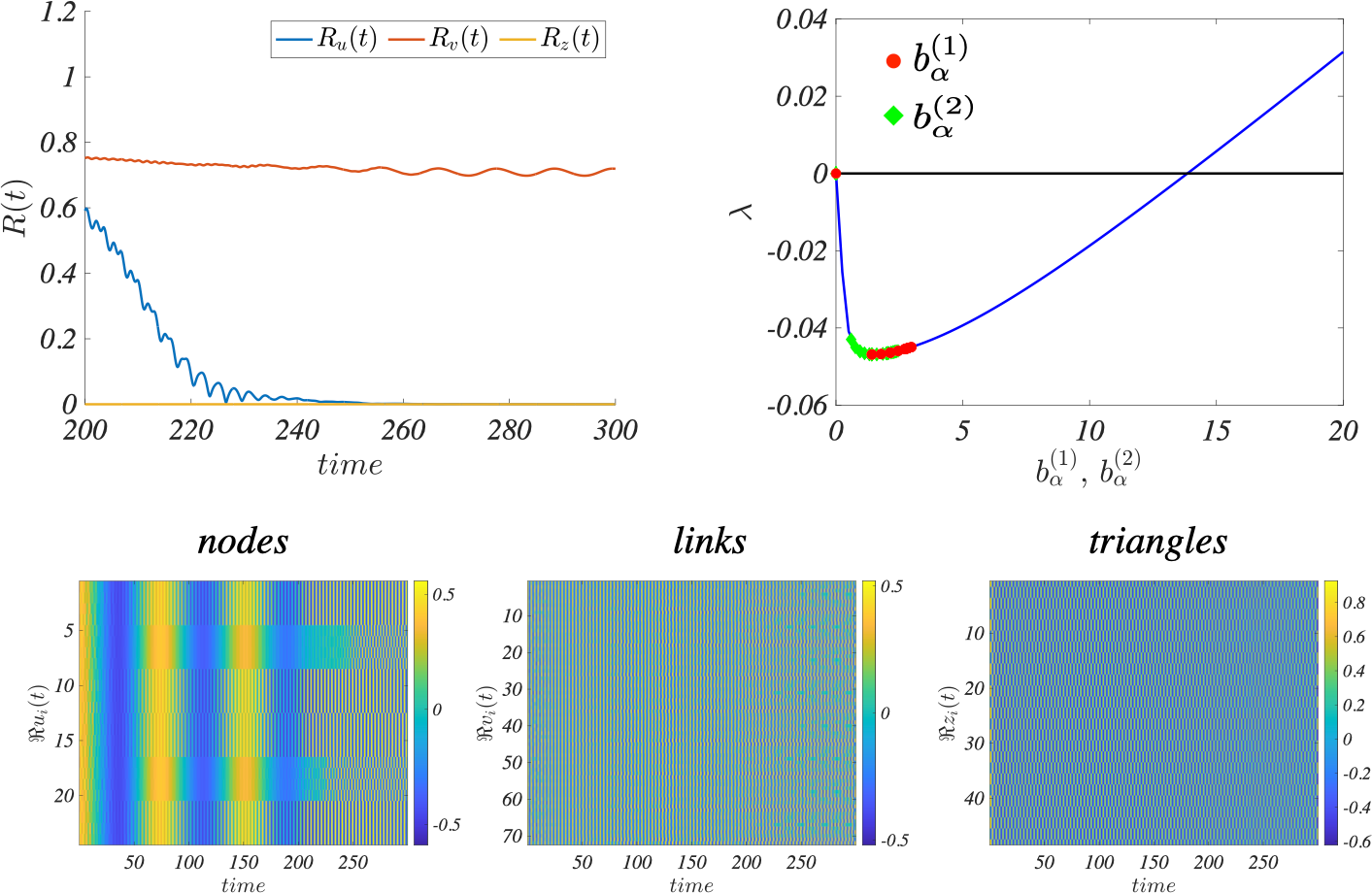}
\caption{Triangulated unweighted $2$-Torus does not supporting Global Topological Synchronization. Top left panel: we show the order parameters for nodes, links and triangles as a function of time.
Top right panel: we report the Master Stability Function, i.e., the maximum of the real part of the characteristic roots, as a function of $b_\alpha^{(1)}$ and $b_\alpha^{(2)}$, the singular values of the matrices $\mathbf{B}_1$ and $\mathbf{B}_2$. The bottom panels show the time evolution of (real part of) the topological signal for nodes, links and triangles. The used parameters are: $\sigma_\Re = 0.2$, $\sigma_\Im = 0.3$, $\beta_\Re = 1.0$, $\beta_\Im = 1.1$, $\mu^{(0)}_\Re = 1.0$, $\mu^{(0)}_\Im = -0.5$, $(\mu_{1}^{(1)})_\Re=(\mu_{1}^{(2)})_\Re=-0.5$, $(\mu_{1}^{(1)})_\Im=(\mu_{1}^{(2)})_\Im=-0.24$, $\mu^{(2)}_\Re = 1.0$, $\mu^{(2)}_\Im = -0.5$.}
\label{fig:TW2TorusNoSynch}
\end{figure}

\begin{figure}[h]
\centering
\includegraphics[width=0.98\linewidth]{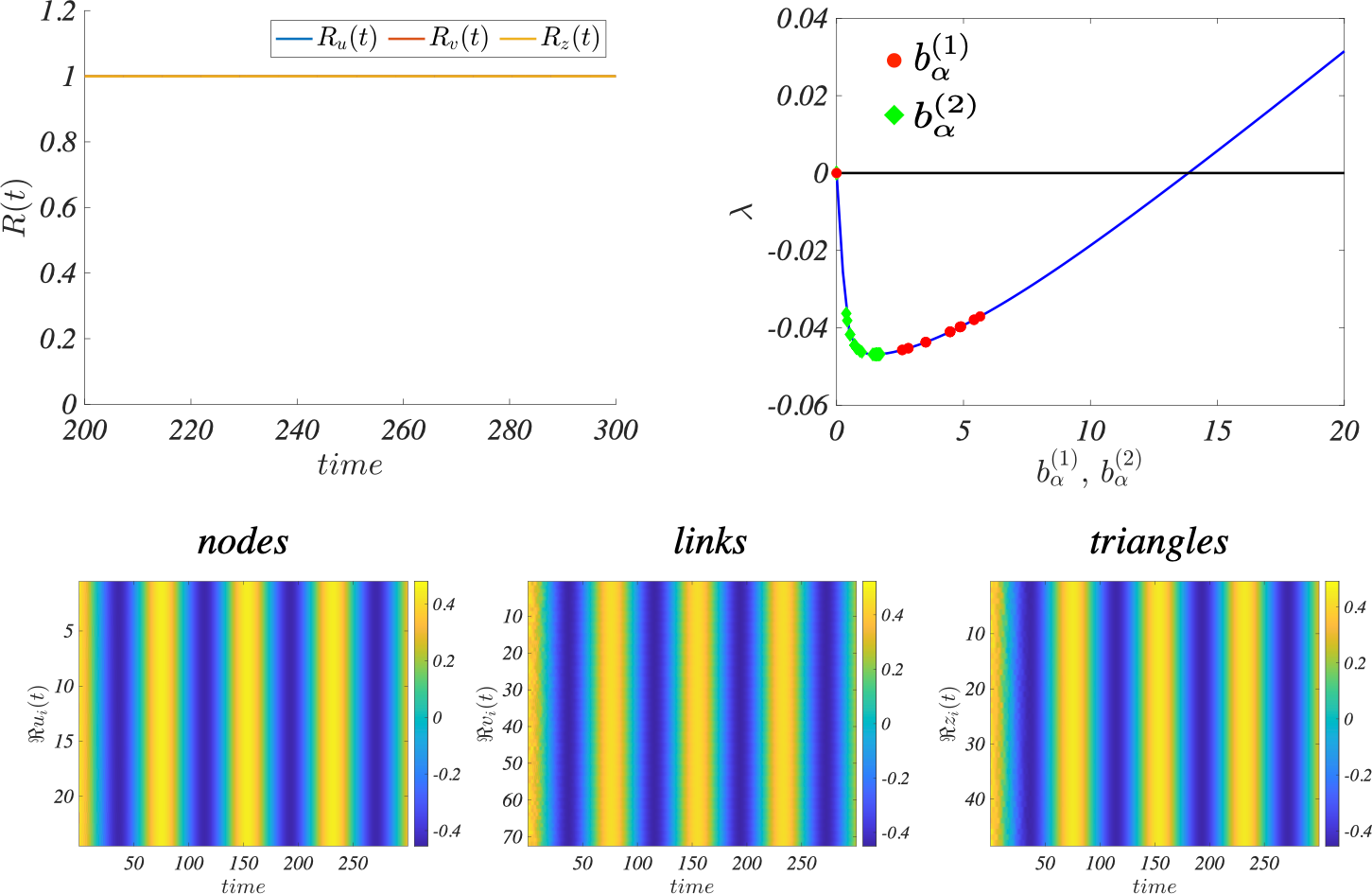}
\caption{Triangulated Weighted $2$-Torus supporting Global Topological Synchronization. Top left panel: we show the order parameters for nodes, links and triangles as a function of time.
Top right panel: we report the Master Stability Function, i.e., the maximum of the real part of the characteristic roots, as a function of $b_\alpha^{(1)}$ and $b_\alpha^{(2)}$, the singular values of the matrices $\mathbf{B}_1$ and $\mathbf{B}_2$. Bottom panels show the time evolution of (real part of) the topological signal for nodes, links and triangles. The used parameters are: $\sigma_\Re = 0.2$, $\sigma_\Im = 0.3$, $\beta_\Re = 1.0$, $\beta_\Im = 1.1$, $\mu^{(0)}_\Re = 1.0$, $\mu^{(0)}_\Im = -1.5$, $(\mu_{1}^{(1)})_\Re=(\mu_{1}^{(2)})_\Re=-0.5$, $(\mu_{1}^{(1)})_\Im=(\mu_{1}^{(2)})_\Im=-0.24$, $\mu^{(2)}_\Re = 1.0$, $\mu^{(2)}_\Im = -1.5$. The edges weights have been set to $w_1=w_2=4$, $w_3=1$ (see Fig.~\ref{fig:2Dtorus} for the convention about the weights definition).}
\label{fig:TW2TorusSynch}
\end{figure}

In order to provide evidence for these statements, we have considered a Weighted Triangulated Torus and its unweighted version whose definition and spectral properties are discussed in~\ref{ApA}.
In Figure~\ref{fig:TW2TorusNoSynch}, we show evidence of lack of GTDS on the unweighted version of the triangulated torus, while in Figure~\ref{fig:TW2TorusSynch}, we demonstrate that with a suitable definition of edges weights and dynamical parameters, GTDS can be achieved. 


\section{Conclusions}

In this work we have combined algebraic topology with nonlinear dynamics to define and fully investigate Global Topological Dirac Synchronization (GTDS). This novel dynamical state of simplicial and cell complexes occurs when all the topological signals defined on any simplex or cell of the higher-order networks are inter-dimensionally coupled via the Dirac operator and obey the same dynamics.
We have  developed a general theory for studying GTDS, by investigating the topological conditions for the existence of this state and the dynamical conditions for its stability. For ease of notation, we have focused  on a $K=2$ cell complex. However, this approach is readily generalizable to cell complexes of arbitrary dimension $K$.
On a $1$-dimensional simplicial complex (i.e., a network), this state exists as long as the network is Eulerian and the dynamics of the uncoupled system is stable and, thus, can be observed as long as  suitable dynamical conditions are met. On the $2$-dimensional case, however, this dynamical state is more rare. The $K$-dimensional square lattice tessellation of the torus is here shown to allow for GTDS. However, $K=2$ dimensional unweighted simplicial complexes can never sustain GTDS. For $K=2$ dimensional simplicial complexes to be able to sustain a GTDS suitable weights need to be chosen.
Our results are discussed by considering a Stuart-Landau dynamics for the topological signals and by studying the stability of the GDTS with advanced dynamical systems and hyperparameters optimization techniques, developed in  machine learning literature \cite{liaw2018tune}. 
Our results go beyond the specific case considered, not only in terms of the topology, as discussed above, but also regarding the dynamics. {  First of all, the Master Stability Function formalism is straightforwardly extended to dynamical systems of $n$ dimensions, including also chaotic dynamics. Then, the behavior of the Stuart-Landau model is representative for all systems with oscillatory behavior. In fact, the Stuart-Landau is the normal form of the supercritical Hopf-Andronov bifurcation, meaning any system exhibiting a stable limit cycle can be reduced to the Stuart-Landau through a technique called center manifold reduction \cite{kuramoto2019concept,Kuramoto}. Lastly, any system with periodic behavior, including thus the Stuart-Landau model, can be reduced to a phase description through another reduction technique, called phase reduction \cite{nakao2016phase}, from which one obtains Kuramoto-like models, extensively studied in the case of topological signals \cite{millan2020explosive}, also including the Dirac coupling \cite{calmon2023local}.} Given the plethora of applications of chaotic synchronization, limit cycles and phase models, these results open new perspectives on the theory of synchronization phenomena occurring on higher-order networks.
{  The GTDS state greatly generalizes the global synchronize state on graph and networks that has found many applications in natural and technological systems and as such could find applications ranging from biological rhythms to power-grids. Moreover, the  GTDS state proposed and investigated here paves the way for experimental realization  of this new dynamical state in future technologies such as  nano-oscillators.}

\section*{Supplementary Movie}
See supplementary movie S1 to appreciate the temporal evolution of the topological signals on nodes, links and faces, toward global synchronisation for the Square Lattice Tessellation of the 2D Torus (SLTT) with the same parameters as in Figure 5.

\section*{Acknowledgements} The authors (T.C. and G.B.) would like to thank the Isaac Newton Institute for Mathematical Sciences, Cambridge, for support and hospitality during the programme Hypergraphs: Theory and Applications, where work on this paper was undertaken. R.M. is supported by a JSPS postdoctoral fellowship, grant 24KF0211. This work was supported by EPSRC grant EP/R014604/1 (T. C. and G. B.)  partially supported by a grant from the Simons Foundation (G. B.).

\section*{Code availability.}
The codes used in this work are freely available at the repository \cite{github_repo}.

\appendix
\section{Proofs of Eq. (\ref{decPsi}) and Eq.(\ref{31}).}
\label{ApA}

	In order to prove Eq.(\ref{decPsi}) let us apply 
	 the projector operator ${\bm \Pi}_{[k]}={\bf P}_{[k]}\otimes {\bf I}_d$ where ${\bf P}_{[k]}$ is defined in Eq. (\ref{projectors}) to both sides of Eq. (\ref{deltaPsi}) by obtaining
	\be
	{\bm \Pi}_{[k]}\frac{ d\delta{\bf X}}{dt}=\frac{d\delta{\bf X}_{[k]}}{dt}={{\bm \Pi}}_{[k]}{\boldsymbol{\mathcal{J}_f}}\delta{\bm \Psi}-{{\bm \Pi}}_{[k]}\Dirac{\boldsymbol{\mathcal{J}_h}}\delta{\bf X}\, .\label{dPsin}
	\ee
	We observe that  \be {{\bm \Pi}}_{[k]}\Dirac=\Dirac{{\bm \Pi}}_{[k]}=\bm\gamma_{[k]}{\boldsymbol{\mathcal{D}}}_{[k]}{{\bm \Pi}}_{[k]}\, ,
 \ee 
 hence we can write Eq. (\ref{dPsin}) as 
	\be
	\frac{{d\delta {\bf X}}_{[k]}}{dt}={{\bm \Pi}}_{[k]}{\boldsymbol{\mathcal{J}}_{\bf f}}\delta{\bm \Psi}-\bm\gamma_{[k]}{\boldsymbol{\mathcal{D}}}_{[k]}{{\bm \Pi}}_{[k]}{\boldsymbol{\mathcal{J}}_{\bf h}}\delta{\bf X}\, .\label{dPsi_1}
	\ee
	In order to simplify this equation let us  note  that \be
	{{\bm \Pi}}_{[k]}={\boldsymbol{\mathcal{L}}}_{[k]}{\boldsymbol{\mathcal{L}}}_{[k]}^{+}\otimes {\bf I}_{N_k} \ee
	and since ${\boldsymbol{\mathcal{L}}}_{[k]}$ is block diagonal then  ${{\bm \Pi}}_{[k]}$ commutes with  ${\boldsymbol{\mathcal{J}}}_{\bf f}$.
	 We thus obtain
	\be
	{{\bm \Pi}}_{[k]}{\boldsymbol{\mathcal{J}}_{\bf f}}{\bf X}={\boldsymbol{\mathcal{J}}_{\bf f}}{{\bm \Pi}}_{[k]}{\bf X}={\boldsymbol{\mathcal{J}}_{\bf f}}{{\bf X}}_{[k]}={\boldsymbol{\mathcal{J}}_{\bf f}^{[k]}}{{\bf X}}_{[k]}\, .
	\ee
	and similarly
	\be
	{{\bm \Pi}}_{[k]}{\boldsymbol{\mathcal{J}_h}}{\bf X}={\boldsymbol{\mathcal{J}}_{\bf f}}{{\bm \Pi}}_{[k]}{\bf X}={\boldsymbol{\mathcal{J}}_{\bf h}}{\bf X}_{[k]}={\boldsymbol{\mathcal{J}}_{\bf h}^{[k]}}{{\bf X}}_{[k]}\, .
	\ee
	that proves  Eq.(\ref{decPsi}) that we rewrite here for convenience
 \be
 \frac{{d\delta {\bf X}}_{[k]}}{dt}={\boldsymbol{\mathcal{J}}_{\bf f}}^{[k]}\delta{\bf X}_{[k]}-\bm\gamma_{[k]}{\boldsymbol{\mathcal{D}}}_{[k]}{\boldsymbol{\mathcal{J}}_{\bf h}}^{[k]}\delta{\bf X}_{[k]}\, .
 \ee
In order to prove Eq.(\ref{31}) let us notice that the harmonic component of the variation $\delta {\bf X}$ is given by 
 \be
 \delta {\bf X}^{\textrm{harm}}=({\bf I}_{d\mathcal{N}}-{\bm \Pi}_{[1]}-{\bm \Pi}_{[2]})\delta{\bf X}\, ,
 \ee
Since $({\bf I}_{d\mathcal{N}}-{\bm \Pi}_{[1]}-{\bm \Pi}_{[2]})\Dirac=0$, starting from Eq. (\ref{deltaPsi}) we obtain Eq.(\ref{31}), i.e.
\be
\frac{\delta {\bf X}^{\textrm{harm}}}{dt}={\boldsymbol{\mathcal{J}}_{\bf f}}{\bf X}^{\textrm{harm}}\, .
\ee

 { 
\section{Proof of Proposition~\ref{prop:coeffb2}}
\label{sec:profprop}

The aim of this Section is to prove Proposition~\ref{prop:coeffb2} and how its application can be used to prove that the stability of the GTDS depends on $b_\alpha^2$.

Let us recall here the Proposition~\ref{prop:coeffb2}
\begin{proposition}
Let us consider a square matrix of the form
\begin{equation}
\label{eq:matM}
 \mathbf{M}(w)= \left(
\begin{matrix}
 \mathbf{A}_1 & w   \mathbf{A}_2\\
 w \mathbf{A}_3 &  \mathbf{A}_4
\end{matrix}
 \right)
\end{equation}
where $ \mathbf{A}_i$, $i=1,2,3,4$, are four generic square matrices and $x$ a real parameter. Then for any integer $k$ we have
\begin{equation}
\label{eq:matMk}
 \mathbf{M}^{2k}(w)= \left(
\begin{matrix}
 p^{(k)}_1(w^2) & w p^{(k)}_2(w^2)\\
 w p^{(k)}_3(w^2) &  p^{(k)}_4(w^2)
\end{matrix}
 \right)
\end{equation}
where $p_1^{(k)}(t)$ and $p_4^{(k)}(t)$, resp. $p_2^{(k)}(t)$ and $p_3^{(k)}(t)$, are polynomials of degree $k$, resp. $k-1$, in the variable $t$, with matrix coefficients depending on the matrices $\mathbf{A}_i$.
\end{proposition}

\noindent{\textit{Proof}.}
The proof can be done by recurrence on the integer $k$. Let $k=1$, then
\begin{equation*}
 \mathbf{M}^2(w)=\left(
\begin{matrix}
 \mathbf{A}_1^2+w^2 \mathbf{A}_2\mathbf{A}_3& w  (\mathbf{A}_1 \mathbf{A}_2+\mathbf{A}_2 \mathbf{A}_4)\\
 w (\mathbf{A}_3 \mathbf{A}_1+\mathbf{A}_4 \mathbf{A}_3)&  w^2\mathbf{A}_2\mathbf{A}_2+\mathbf{A}_4^2
\end{matrix}
 \right)\equiv
 \left(
\begin{matrix}
p_1^{(1)}(w^2)& w p_2^{(1)}\\
 w p_3^{(1)}&  p_4^{(1)}(w^2)
\end{matrix}
 \right)\, ,
\end{equation*}
where the polynomials $p_i^{(1)}$, $i=1,\dots,4$, are defined by the last equality. Moreover $p_1^{(1)}(t)$ and $p_4^{(1)}(t)$ are of degree $1$ in the variable $t$, while $p_2^{(1)}(t)$ and $p_3^{(1)}(t)$ are constant, namely of degree $0$ in the variable $t$.

Let us assume Eq.~\eqref{eq:matMk} to hold true for all $k\leq m$ and let us prove its validity for $k=m+1$. Let us thus compute
\begin{eqnarray*}
\hspace*{-15mm}
& & \mathbf{M}^{2(m+1)}(w)=  \mathbf{M}^{2}(w) \mathbf{M}^{2m}(w)=\left(
\begin{matrix}
 p^{(1)}_1(w^2) & w p^{(1)}_2(w^2)\\
 w p^{(1)}_3(w^2) &  p^{(1)}_4(w^2)
\end{matrix}
 \right)\left(
\begin{matrix}
 p^{(m)}_1(w^2) & w p^{(m)}_2(w^2)\\
 w p^{(m)}_3(w^2) &  p^{(m)}_4(w^2)
\end{matrix}
 \right)\notag\\
\hspace*{-15mm} &=&\left(
\begin{matrix}
 p^{(1)}_1(w^2) p^{(m)}_1(w^2)+w^2 p^{(1)}_2(w^2)p^{(m)}_3(w^2) & w (p^{(1)}_1(w^2)p^{(m)}_2(w^2)+p^{(1)}_2(w^2)p^{(m)}_4(w^2))\\
w (p^{(1)}_3(w^2)p^{(m)}_1(w^2)+p^{(1)}_4(w^2)p^{(m)}_3(w^2)) &   p^{(1)}_4(w^2) p^{(m)}_4(w^2)+w^2 p^{(1)}_3(w^2)p^{(m)}_2(w^2) 
\end{matrix}
 \right)\notag\\
\hspace*{-15mm} &\equiv &\left(
\begin{matrix}
 p^{(m+1)}_1(w^2) & w p^{(m+1)}_2(w^2)\\
 w p^{(m+1)}_3(w^2) &  p^{(m+1)}_4(w^2)
\end{matrix}
 \right)\, .
\end{eqnarray*}
The polynomials $p^{(m+1)}_i$, $i=1,\dots,4$, are defined by the last equality and one can prove by direct inspection that they depend on $w^2$. Moreover
\begin{eqnarray*}
\hspace*{-15mm}\deg  p^{(m+1)}_1(t) &=&\max \{ \deg   p^{(1)}_1(t) + \deg p^{(m)}_1(t), \deg p^{(1)}_2(t) + \deg p^{(m)}_3(t)+1\}=m+1\\
\hspace*{-15mm}\deg  p^{(m+1)}_2(t) &=&\max \{ \deg   p^{(1)}_1(t) + \deg p^{(m)}_2(t), \deg p^{(1)}_2(t) + \deg p^{(m)}_4(t)\}=m\\
\hspace*{-15mm}\deg  p^{(m+1)}_3(t) &=&\max \{ \deg   p^{(1)}_3(t) + \deg p^{(m)}_1(t), \deg p^{(1)}_4(t) + \deg p^{(m)}_3(t)\}=m\\
\hspace*{-15mm}\deg  p^{(m+1)}_4(t) &=&\max \{ \deg   p^{(1)}_4(t) + \deg p^{(m)}_4(t), \deg p^{(1)}_3(t) + \deg p^{(m)}_2(t)+1\}=m+1\, ,
\end{eqnarray*}
and this concludes the proof.\\

A similar conclusion can be proved for $\mathbf{M}^{2k+1}(w)$.\\

We can now prove that the stability of GTDS depends on $b_\alpha^2$. For this we have to compute the spectrum of the matrix $\mathbf{M}(b_{\alpha})$ given by~\eqref{eq:DiraccouplinprojMat}, which is of the form~\eqref{eq:matM}. The eigenvalues are the root of the characteristic polynomial $\det (\mathbf{M}(b_{\alpha}) - \lambda\mathbf{I})=0$. The Cayley–Hamilton theorem allows to express the determinant of a matrix as linear combination of the trace of the powers of such matrix, hence the latter determinant is obtained as a linear combination of the trace of the powers of $\mathbf{M}(b_{\alpha})$. By using the Proposition just proven, the latter are polynomials of $b_{\alpha}^2$ and this conclude the claim.}

\section{GTDS regions through the Routh-Hurwitz stability criterion}
\label{sec:ApC}
\setcounter{figure}{0}

As reported in the Main Text, the conditions to obtain Global Topological  Dirac Synchronization (GTDS) can be obtained by studying the stability of the polynomial $p_\alpha(\lambda)$ through the Routh-Hurwitz criterion. More precisely, there is a necessary condition, also known in the literature as Stodola criterion~\cite{Bissell}, which tells us that the roots of $p_\alpha(\lambda)$ have negative real part if all the coefficients are positive:
\begin{equation}
\label{eq:RHnec}
a_0>0 \, ,a_1>0\, ,a_2>0\,,a_3>0 \text{ and }a_4>0\, ,
\end{equation}
while a sufficient condition is
\begin{eqnarray}
\label{eq:RHsuff}
a_0>0 \, ,a_1>0\, ,a_1 a_2-a_3a_0>0\,,\notag\\a_3(a_1a_2-a_3a_0)-a_4a_1^2>0 \text{ and }a_4>0\, .
\end{eqnarray}
From the explicit form of the coefficients, given by Eq. \eqref{eq:coeff}, we have that $a_0>0$ and $a_1>0$. The third condition, i.e., $ a_1 a_2-a_3a_0>0$, gives \begin{equation}
16\sigma_\Re^3 + \left(\frac{4b_\alpha^2 \beta_\Im \Im(\mu^{(0)} \mu^{(1)})}{\beta_\Re} - 4b_\alpha^2 \Re(\mu^{(0)} \mu^{(1)})\right) \sigma_\Re > 0,
\end{equation} which could be, in principle, treated analytically together with the condition $a_4>0$. However, the fourth condition, i.e., $a_3(a_1a_2-a_3a_0)-a_4a_1^2>0$, gives a cumbersome expression even in the case in which the coefficients $\sigma$ and $\beta$ are the same for the two dynamical system defined on nodes and edges, thus the problem needs to be solved numerically. This would give us the conditions for the Master Stability Function (MSF) to be negative and, hence, achieve GTDS. Nonetheless, as pointed out in the Main Text, the above conditions are, in our case, sufficient but not necessary. In fact, given the discrete nature of the spectra of the involved operators, we do not need the MSF to be always negative to deal with GTDS, as long as it is negative in correspondence of the discrete eigenvalues of the operators $B_k$.

In conclusion, the Routh-Hurwitz criterion is not convenient to use in this context, because the conditions are too restrictive and, even though we can obtain the parameter regions in which the system exhibits GTDS, those regions are likely to be a small subset of all the possible configurations in which such a dynamics can be achieved. However, let us point out that the Routh-Hurwitz criterion can be useful once the model parameters are different for the systems on different simplexes. In such a case, GTDS can still be achieved as long as the parameters are such that the frequencies of the systems are the same for every simplex, but the method of the Main Text can be extremely cumbersome due to the many different parameters in place. Then, the Routh-Hurwitz criterion can be a good starting point for finding the GTDS regions where the MSF is always negative. Such regions can then be further extended by exploring the neighborhoods of the boundaries.

{ 
\section{GTDS regions through the Master Stability Function}
\label{sec:perturbcalc}

The aim of this section is to provide some detail about the formula~\eqref{eq:rootapprox} returning the asymptotic development of the eigenvalues, i.e., the roots of the characteristic polynomial~\eqref{eq:polcar}, of the linearized system~\eqref{eq:maineqDiracSLlinproj} in the limit $b_\alpha\rightarrow 0$.

Because we are perturbing the limit cycle of the SL model and because the dynamics on links and node decouple once we set $b_1=0$, it is clear that the roots of the characteristic polynomials are given by
\begin{equation*}
 \lambda_1(b_1)= \lambda_2(b_1)=-2\sigma_\Re\quad \text{ and }\quad \lambda_3(b_1)= \lambda_4(b_1)=0\, .
\end{equation*}

Les us now consider small but non zero $|b_\alpha|$ and let us analyze the asymptotic expression of the roots; it is clear that $\lambda_1(b_\alpha)$ and $\lambda_2(b_\alpha)$ will (generically) assume different negative values if $|b_\alpha|$ is small enough. On the other hand, the vanishing roots, $\lambda_3(b_\alpha)$ and $\lambda_4(b_\alpha)$, can bifurcate either by remaining real numbers and assume positive or negative values, or can become complex conjugate numbers with positive or negative real part. 

We can thus look for an expansion of $\lambda_3(b_\alpha)$ and $\lambda_4(b_\alpha)$ of the form $\lambda_j(b_\alpha)=\lambda_j^{(1)} b_\alpha+\lambda_j^{(2)} b_\alpha^2+\dots$, for $j=3,4$ and for small $b_\alpha$. The unknown values $\lambda_j^{(1)}$ and $\lambda_j^{(2)}$ can be determined by inserting this ansatz into the characteristic polynomial~\eqref{eq:polcar} and equate terms with the same powers of $b_\alpha$. More precisely, let us rewrite for reading ease, the characteristic polynomial
\begin{equation*}
p_\alpha(\lambda) = \det \left(\mathbf{J}_\alpha-\lambda \mathbf{I}_4\right)=a_0\lambda^4+a_1 \lambda^3+a_2 \lambda^2+a_3\lambda + a_4\, ,
\end{equation*}
where the coefficients $a_j$, $j=0,\dots,4$ are given by~\eqref{eq:coeff}, also repeated here:
\begin{eqnarray*}
a_0 &=& 1,\nonumber \\
a_1 &=& -2\mathrm{tr}\left(\mathbf{J}\right) = 4\sigma_\Re,\nonumber \\
a_2 &=& - 2b_\alpha^2 (\mu^{(0)} \mu^{(1)})_\Re + 4 (\sigma_\Re)^2=:b_\alpha^2 a_2^{(2)}+a_2^{(0)}, \nonumber\\
a_3 &=& -4 b_{\alpha}^2 \frac{\sigma_\Re}{\beta_\Re} \left(\beta_\Re (\mu^{(0)} \mu^{(1)})_\Re + \beta_\Im (\mu^{(0)} \mu^{(1)})_\Im\right)=:b_\alpha^2 a_3^{(2)}, \\
a_4 &=& b_\alpha^4|\mu^{(0)}_\Re|^2 |\mu^{(1)}|^2  -4b_\alpha^2 \frac{\sigma_\Re^2}{\beta_\Re^2}\left(\beta_\Re^2\mu^{(0)}_\Re\mu^{(1)}_\Re +\beta_\Im\beta_\Re\mu^{(0)}_\Re\mu^{(1)}_\Im \right. \nonumber \\
    && \left. +\beta_\Im\beta_\Re\mu^{(0)}_\Im\mu^{(1)}_\Re +\beta_\Im^2\mu^{(0)}_\Im\mu^{(1)}_\Im\right)=:b_\alpha^4 a_4^{(4)}+b_\alpha^2 a_4^{(2)},\nonumber
\end{eqnarray*}
where we also defined $a_i^{(k)}$ to be the coefficient of $b_\alpha^k$ in the polynomial coefficient $a_i$, with $i=2,3,4$.

Proceeding as stated above, we get:
\begin{eqnarray*}
\hspace*{-15mm}& &0= p_\alpha(\lambda_j(b_\alpha)) =a_0b_\alpha^4\left(\lambda_j^{(1)} +\lambda_j^{(2)} b_\alpha+\dots\right)^4+a_1 b_\alpha^3\left(\lambda_j^{(1)} +\lambda_j^{(2)} b_\alpha+\dots\right)^3+\\
\hspace*{-15mm} &+&\left(b_\alpha^2 a_2^{(2)}+a_2^{(0)}\right) b_\alpha^2\left(\lambda_j^{(1)} +\lambda_j^{(2)} b_\alpha+\dots\right)^2+b_\alpha^2 a_3^{(2)} b_\alpha \left(\lambda_j^{(1)} +\lambda_j^{(2)} b_\alpha+\dots\right) + b_\alpha^4 a_4^{(4)}+b_\alpha^2 a_4^{(2)}\, ,
\end{eqnarray*}
and expanding into powers of $b_\alpha$ we get
\begin{equation*}
 0=a_4^{(2)}+a_2^{(0)}\left[\lambda_j^{(1)}\right]^2 + b_\alpha\Big\{a_1\left[\lambda_j^{(1)}\right]^3+2a_2^{(0)}\lambda_j^{(1)}\lambda_j^{(2)}+a_3^{(2)}\lambda_j^{(1)}\Big\}+\mathcal{O}(b_\alpha^2)\, .
\end{equation*}
To satisfy the latter equation for all (small enough) $|b_\alpha|$ we must impose
\begin{eqnarray}
 a_4^{(2)}+a_2^{(0)}\left[\lambda_j^{(1)}\right]^2=0\label{eq:constr1}\\
 a_1\left[\lambda_j^{(1)}\right]^3+2a_2^{(0)}\lambda_j^{(1)}\lambda_j^{(2)}+a_3^{(2)}\lambda_j^{(1)} = 0\label{eq:constr2}\, .
\end{eqnarray}
The first equation returns
\begin{equation*}
 \left[\lambda_j^{(1)}\right]^2 = -\frac{a_4^{(2)}}{a_2^{(0)}}=\frac{C_1}{\beta_\Re^2}\, ,
\end{equation*}
where we used the explicit expression for the coefficients $a_i^{(k)}$ given above and we defined
\begin{equation*}
 C_1=\beta_\Re^2\mu^{(0)}_\Re\mu^{(1)}_\Re +\beta_\Im\beta_\Re\mu^{(0)}_\Re\mu^{(1)}_\Im +\beta_\Im\beta_\Re\mu^{(0)}_\Im\mu^{(1)}_\Re +\beta_\Im^2\mu^{(0)}_\Im\mu^{(1)}_\Im\equiv (\beta_\Re \mu^{(0)}_\Re+\beta_\Im \mu^{(0)}_\Im)(\beta_\Re \mu^{(1)}_\Re+\beta_\Im \mu^{(1)}_\Im)\, .
\end{equation*}

The second equation~\eqref{eq:constr2} provides
\begin{equation*}
 \lambda_j^{(2)}=-\frac{a_1}{2a_2^{(0)}} \left[\lambda_j^{(1)}\right]^2-\frac{a_3^{(2)}}{2a_2^{(0)}}=-\frac{1}{2\sigma_\Re} \frac{C_1}{\beta_\Re^2}+\frac{1}{2\sigma_\Re\beta_\Re} \left(\beta_\Re (\mu^{(0)} \mu^{(1)})_\Re + \beta_\Im (\mu^{(0)} \mu^{(1)})_\Im\right)\, ,
\end{equation*}
once we used the already computed value for $\lambda_j^{(1)}$ and the expressions of the coefficients. 

In conclusion we have proved that for small enough $|b_\alpha|$, the roots $\lambda_3(b_\alpha)$ and $\lambda_4(b_\alpha)$ exhibit the following behavior
\begin{eqnarray*}
\lambda_3(b_\alpha)=  -\frac{\sqrt{C_1}}{\beta_\Re}b_\alpha-\frac{\mu^{(0)}_\Im \mu^{(1)}_\Im}{2\beta_\Re^2\sigma_\Re}|\beta|^2b_\alpha^2+\mathcal{O}(b_\alpha^3) \\
\lambda_4(b_\alpha)=  \frac{\sqrt{C_1}}{\beta_\Re}b_\alpha-\frac{\mu^{(0)}_\Im \mu^{(1)}_\Im}{2\beta_\Re^2\sigma_\Re}|\beta|^2b_\alpha^2+\mathcal{O}(b_\alpha^3)\, ,
\end{eqnarray*}
namely Eqs.~\eqref{eq:rootapprox}.
}

 \section{Spectral properties of the Weighted Triangulated Torus and the Square Lattice Tesselation of the Torus}
 In this appendix we discuss the spectral properties of the two considered $K=2$ dimensional cell complexes that can sustain GTDS:  The Square Lattice Tessellation of the Torus (SLTT) and the  Weighted Triangulated Torus (WTT). The spectrum of the Hodge Laplacians for these latter $2$-dimensional simplicial  complexes has been recently derived in Ref.~\cite{WMCB2024}. 
 
\begin{figure}[h]
\centering
\includegraphics[width=0.9\linewidth]{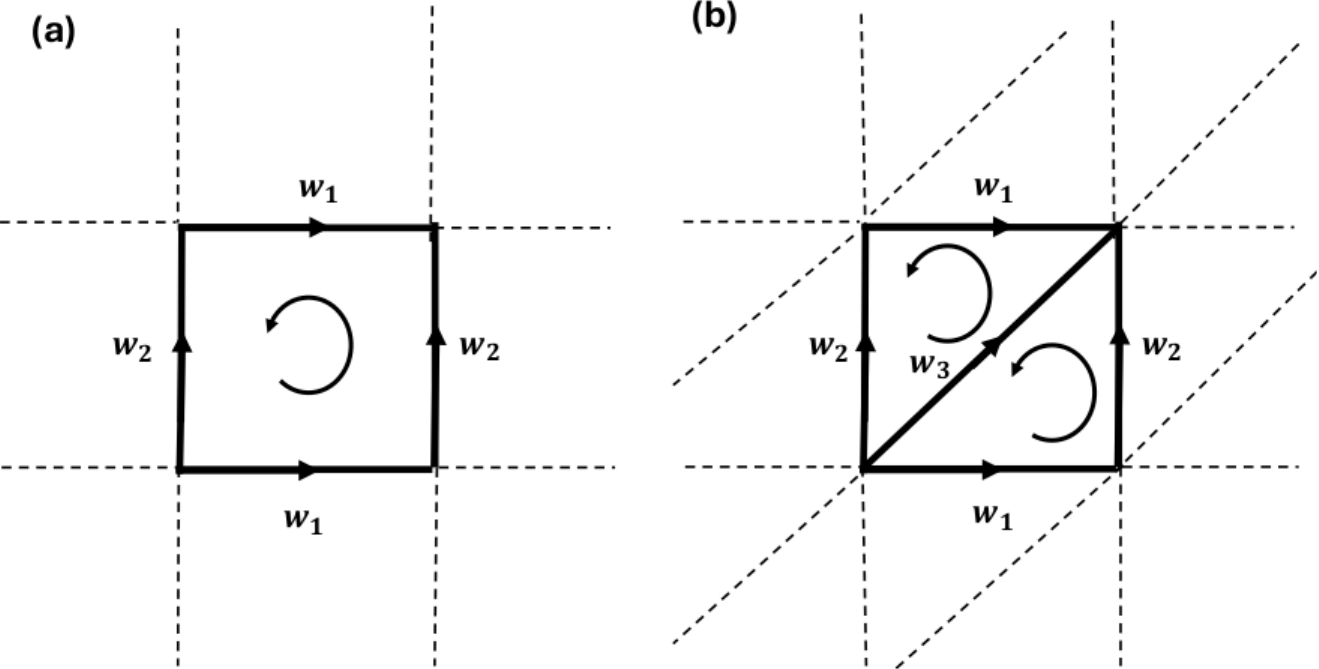} 
\caption{Panel (a): The weighted Square Lattice Tessellation of the  $2$-Torus (SLTT). We report one basic cell with the oriented  edges. The orientation of the square  is shown by using the curved  arrow. The unweighted  square tessellated $2$-torus is obtained by putting $w_1=w_2=1$. Panel (b): The Weighted Triangulated $2$-Torus (WTT). We report one basic cell with the oriented weighted edges. The orientation of each triangle is shown by using the curved  arrow.}
\label{fig:2Dtorus}
\end{figure} 
The SLTT is a $2$-dimensional cell complexes, whose skeleton is a square lattice of size $\hat{L}$ with periodic boundary conditions.
 The WTT is a $2$-dimensional simplicial complex whose skeleton is obtained by using a square lattice of size $\hat{L}$ with periodic boundary conditions where each square is divided into two triangles. Therefore the network skeleton of the WTT is a  regular lattice in which some nodes have degree $6$ and and others degree $4$. 
 
 The orientation of the edges and of the $2$-cells, i.e., squares for the SLTT and triangles for the WTT  are taken according to the convention defined in  Fig.~\ref{fig:2Dtorus}. 
 We consider always non trivial metric matrices ${\bf G}_{1}$ whose diagonal terms are determined by the weights of the edges, while we take always trivial metric matrices on nodes and $2$-cells, i.e. ${\bf G}_0={\bf I}_{N_0}$ and ${\bf G}_2={\bf I}_{N_2}$. 
The weights of the edges are taken in the following way. Horizontal edges have weight $w_1>0$, vertical ones have weight $w_2>0$, while the diagonal ones (for the WTT) have weight $w_3>0$ (see Fig.~\ref{fig:2Dtorus}).

Let us start discussing the spectrum of the  SLTT which is much simpler than the spectrum of the WTT. On a weighted SLTT the eigenvalues  $\Lambda_0^{{(\alpha)}}$ of ${\bf L}_0$ are given by 
\be
\Lambda^{{(\alpha)}}_{0} = 4w_1 \sin^2\left(\frac{k_x}{2}\right) + 4w_2 \sin^2\left(\frac{k_y}{2}\right)
\label{L2la}
\ee
while the eigenvalues $\Lambda_2^{{(\alpha)}}$ of ${\bf L}_2$ are given by 
\be
\Lambda^{{(\alpha)}}_{2} = 4\frac{1}{w_2} \sin^2\left(\frac{k_x}{2}\right) + 4\frac{1}{w_1} \sin^2\left(\frac{k_y}{2}\right),
\label{L2lb}
\ee
where here and in the following we indicate the wave number as $\mathbf{k}=(k_x,k_y)$ with $k_x,k_y$ having values  in the discrete sets $k_x = \frac{2 \pi n_x}{\hat{L}}$ and $k_y = \frac{2\pi n_y}{\hat{L}}$ with with $0\leq n_x<\hat{L}$ and $0\leq n_y<\hat{L}$,  $\hat{L}$ being the number of elementary squares in the $2$-dimensional torus both horizontally and vertically.
Let us observe that in the previous equation with a slight abuse of notation we have indexed the eigenvalues with $\alpha=1,\dots,N_0=L^2$ and equivalently, with the double index $\mathbf{k}=(k_x,k_y)$.
From Eq.(\ref{L2la}) and Eq.(\ref{L2lb}) we observe that  if the SLTT is unweighted, i.e., $w_1=w_2=1$  the spectrum of ${\bf L}_0$ coincides with the spectrum of ${\bf L}_2^{\textrm{down}}$ due to the self-duality of the lattice. 
Importantly, we note  that for this cell complex the conditions Eq.(\ref{conditions}) for the existence of the GTDS are satisfied already for the unweighted SLTT thus justifying our choice to focus on this relevant $K=2$ cell complex.

Let us now focus on the WTT. In order to allow for GTDS, i.e., to satisfy Eq.(\ref{conditions}); as already discussed in Ref.~\cite{WMCB2024}
we must impose that the weights of the WTT defined in Fig. \ref{fig:2Dtorus} satisfy 
\be
\frac{1}{\sqrt{w_1}} + \frac{1}{\sqrt{w_2}}=\frac{1}{\sqrt{w_3}}\, .
\ee 
From the latter it follows that once considering Unweighted Triangulated Tori, i.e., by assuming $w_1=w_2=w_3$, this condition is not met, thus a non trivial choice of the weights is necessary to observe GTDS on WTT.

Under this conditions, the eigenvalues $\Lambda_0^{{(\alpha)}}$ of ${\bf L}_{0}$,  can be explicitly computed~\cite{WMCB2024} and they read:
\begin{equation}
    \label{eq:eigval0TTW}
    \Lambda^{{(\alpha)}}_{0} = 4w_1 \sin^2\left(\frac{k_x}{2}\right) + 4w_2 \sin^2\left(\frac{k_y}{2}\right)+ 4w_3 \sin^2\left(\frac{k_x+k_y}{2}\right)
\end{equation}
  Let us observe that the non-zeros eigenvalues of ${\bf L}_0$ coincide with the non-zeros eigenvalues of ${\bf L}^{\textrm{down}}_1$. 

Similarly,  non-zero eigenvalues of ${\bf L}_{2}^{\textrm{down}}$ coincide with the non-zero eigenvalues of   ${\bf L}_{1}^{\textrm{up}}$, and are given by~\cite{WMCB2024}
\be
    \label{eq:eigval2TTW}
     \Lambda^{{(\alpha)}}_{2}=\frac{1}{w_1} + \frac{1}{w_2} + \frac{1}{w_3} \pm |f({\bf k})|,\ee
     where $|f({\bf k})|$ is given by 
     \be
     |f({\bf k})|=\sqrt{\left(\frac{1}{w_1^2}+\frac{1}{w_2^2}+\frac{1}{w_3^2}\right)+\frac{2}{w_1w_2w_3}\left[w_1\cos(k_x)+w_2\cos(k_y)+w_3\cos(k_x+k_y)\right]}.\nonumber 
\ee

\end{document}